\documentclass[prb,twocolumn,aps,superscriptaddress,showpacs]{revtex4-1}

\usepackage{graphicx}
\usepackage{amssymb}
\usepackage{mathrsfs}
\usepackage[english]{babel}
\usepackage{amstext}
\usepackage{amsmath}
\usepackage{bm}

\usepackage{xcolor}
\usepackage{subfig}
\DeclareRobustCommand{\vect}[1]{\bm{#1}}

{

}
\graphicspath{{./}{../}}
\newcommand{\Tsim}{\ensuremath{0.1}}

\newcommand{\gEigV}{\ensuremath{\vect{\phi}}}
\newcommand{\EigV}{\ensuremath{\vect{\hat{\phi}}}}
\newcommand{\MH}{\ensuremath{\vect{\mathcal H}}}
\newcommand{\MHKPM}{\ensuremath{\widetilde{\vect{\mathcal H}}}}
\newcommand{\randVect}{\ensuremath{\mathbf{u}}}
\newcommand{\ww}{\ensuremath{\omega'}}
\newcommand{\wwp}{\ensuremath{\omega_p}}

\begin{document}

\title{Scaling up the lattice dynamics of amorphous materials by orders of magnitude}

\author{Ivan Kriuchevskyi}
\affiliation{Department of Physics ``A. Pontremoli'', University of Milan, via Celoria 16, 20133 Milan, Italy}

\author{Vladimir V. Palyulin}
\affiliation{Centre for Computational and Data-Intensive Science and Engineering, Skolkovo Institute of Science and Technology, Nobelya Ulitsa 3, Moscow, 121205, Russia}

\author{Rico Milkus}
\affiliation{Department of Chemical Engineering and Biotechnology, University of Cambridge, Cambridge
CB3 0AS, U.K.}

\author{Robert M. Elder}
\affiliation{Polymers Branch, U.S. Army Research Laboratory, Aberdeen Proving Ground, MD, USA}
\affiliation{Bennett Aerospace, Inc., Cary, North Carolina 27518, USA}
\affiliation{Center for Devices and Radiological Health, U.S. Food and Drug Administration, Silver Spring, Maryland 20903, USA}

\author{Timothy W. Sirk}
\affiliation{Polymers Branch, U.S. Army Research Laboratory, Aberdeen Proving Ground, MD, USA}

\author{Alessio Zaccone}
\affiliation{Department of Physics ``A. Pontremoli'', University of Milan, via Celoria 16, 20133 Milan, Italy}
\affiliation{Department of Chemical Engineering and Biotechnology, University of Cambridge, Cambridge
CB3 0AS, U.K.}

\begin{abstract}
We generalise the non-affine theory of viscoelasticity for use with large, well-sampled systems of arbitrary chemical complexity. Having in mind predictions of mechanical and vibrational properties of amorphous systems with atomistic resolution, we propose an extension of the Kernel Polynomial Method (KPM) for the computation of the vibrational density of states (VDOS) and the eigenmodes, including the $\Gamma$-correlator of the affine force-field, which is a key ingredient of lattice-dynamic calculations of viscoelasticity. We show that the results converge well to the solution obtained by direct diagonalization (DD) of the Hessian (dynamical) matrix. As is well known, the DD approach has prohibitively high computational requirements for systems with $N=10^4$ atoms or larger. Instead, the KPM approach developed here allows one to scale up lattice dynamic calculations of real materials up to $10^6$ atoms, with a hugely more favorable (linear) scaling of computation time and memory consumption with $N$.
\end{abstract}

\maketitle

\section{Introduction}
For the case of elasticity of centrosymmetric crystalline solids, Born and Huang developed a theory which can straightforwardly predict and compute the elastic moduli from the atomistic structure \cite{BornHuang}. Unfortunately, the task becomes considerably more complex in the case of amorphous materials which lack atomic-scale centrosymmetry. Only recently it was shown that so-called non-affine corrections to the original Born and Huang approach offer a pathway for the prediction of glass viscoelasticity \cite{Lemaitre2006,Milkus2017,Prediction2018}. These corrections account for additional relaxations of atomic positions in non-centrosymmetric cases and result in an overall softening of a material. In our previous work, we examined the non-affine lattice dynamics theory (NALD) against the results produced by bead-spring MD simulations for the case of polymer glasses and found excellent agreement between the two \cite{stiffness,Prediction2018}. 

However, the application of lattice dynamics calculations into the context of materials science has proven more difficult. Lattice dynamical calculations have been demonstrated as a promising path forward to relate the chemical composition of amorphous materials, as defined through atomistic models, with the full range of frequency-dependent viscoelasticity~\cite{Rutledge1994,ElderZacconeSirk:2019}. Similar success has been achieved for the lattice dynamics of simpler systems, such as monoatomic glasses~\cite{Ruocco1996,Beltukov2016}, granular and jammed systems~\cite{OHern2018,Vitelli2009,Tighe2011,Mizuno2016,Ikeda2020} and topological materials~\cite{Vitelli2016}. Yet, two key bottlenecks remain to be solved before a broad class of ``real'' material compositions can be examined. First, realistic atomistic simulations often require relatively large systems on the order of $10^5$ atoms or more \cite{atomistic1,atomistic2}. A key component of lattice dynamics computations is the analysis of the spectral density of dynamical matrices and their eigenvector characteristics or eigenmode spectrum \cite{Rodney2017,Beltukov2016, Prediction2018}. For this relatively large number of atoms, direct diagonalization (DD) of the Hessian matrices ceases to be a viable method to obtain the eigenfrequencies and eigenmodes, since typically DD becomes prohibitive for $N \geq 10^4$ due mainly to memory requirements. Thus, a method to treat larger systems must be established in order to solve the multi-scale problem in computational materials science. Fortunately, within the framework of the NALD approach, it suffices to get the distributions of eigenvalues of the Hessian matrix, which are directly linked to the vibrational density of states, rather than the exact discrete set of values. This allows the use of approximate approaches for direct computation of the vibrational densities of states (VDOS) as well as a quantity computed from the eigenvectors, the correlator of the affine force $\Gamma$. Second, the original theory~\cite{Lemaitre2006,Milkus2017} was developed for single-mass material models. The chemistry of most solids requires a multi-mass representation, and particularly so for models with atomistic detail. Hence, this paper generalizes the NALD framework for the case of multi-component solids. It will then be shown that a Kernel Polynomial Method, based on Chebyshev approximants of the eigenvectors of the Hessian matrix, can yield not only the VDOS, which was shown in previous work, but also the eigenvector-based quantities that are critical in evaluating the viscoelastic moduli of the material within the NALD approach. 

All in all, the framework allows lattice dynamic calculations of viscoelastic moduli of amorphous solids to be performed, for the first time, on systems larger than $N=10^{5}$. From the analysis of the computational performance, it is clear that calculations for $N= 10^{6}$ are now possible. The proposed framework thus provides a working solution to the problem of bridging time and length-scales in the molecular simulation of materials mechanics.

\begin{figure}[h!]
\centering
\includegraphics[angle=-90,width=0.8\columnwidth]{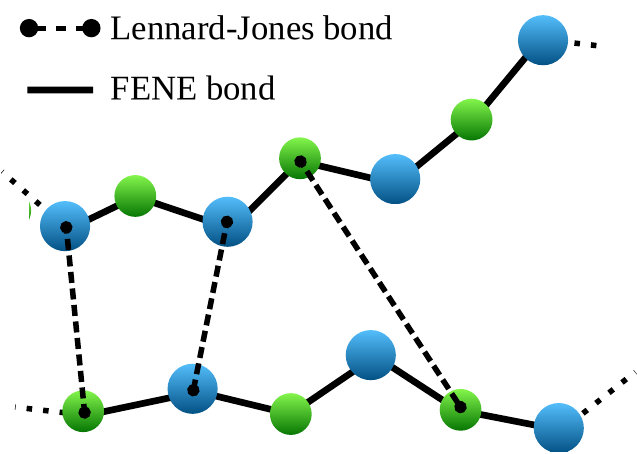}

\includegraphics[width=0.85\columnwidth]{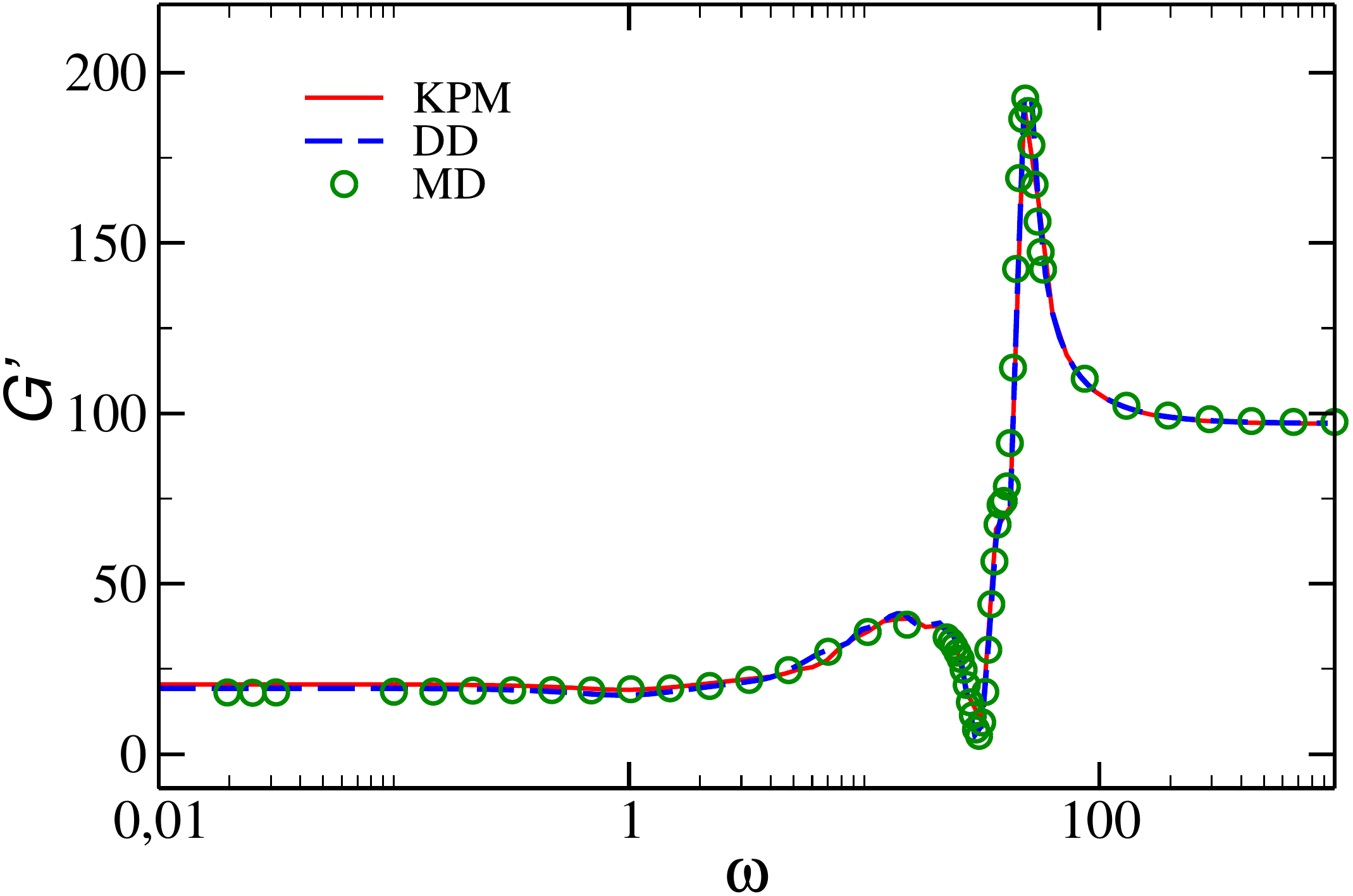}

\includegraphics[width=0.85\columnwidth]{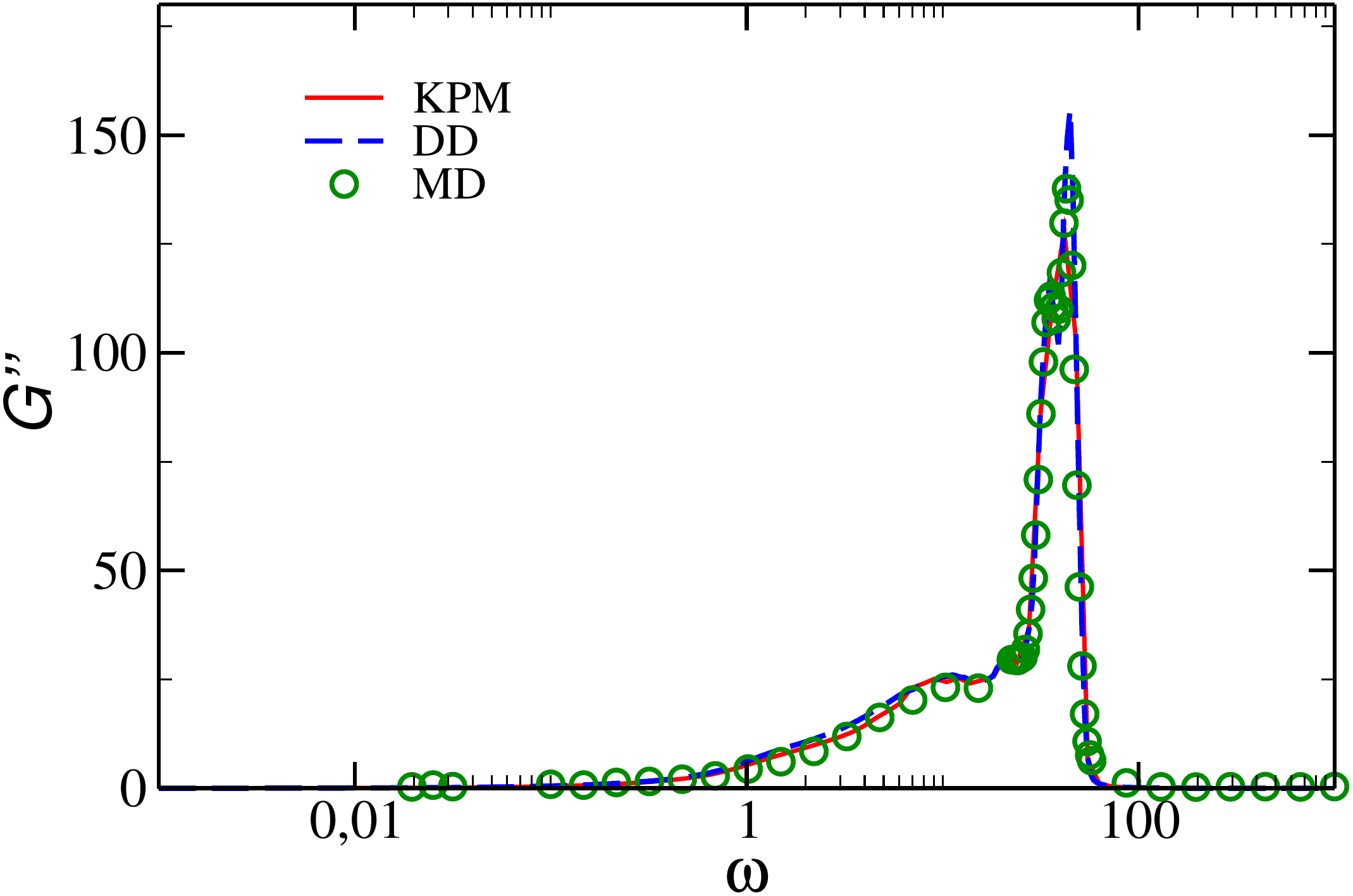}
 \caption{\textit{Panel (a)}: Sketch of two alternating copolymer chains as they appear in the system. The Kremer-Grest model~\cite{Kremer1986} consisting of linear chains of 50 monomers is used. Some of the Lennard-Jones bonds between the chains are depicted as dashed lines. The FENE bonds along the polymer chain are represented as solid lines. The monomers with $m_1=1$ and $m_2=3$ alternate starting from the end of the chain (\textit{AB-configuration}). \textit{Panel~(b-c)}: The values of storage modulus $G'(\omega)$ and loss modulus $G''(\omega)$ respectively, from direct diagonalisation (blue), KPM (red) and MD simulations (black squares) for the polymer system. The well equilibrated glassy system is probed at the temperature $T=0.1 \ll T_\text{g} \approx 0.4$. The total number of monomers in the system is equal to $5 \times 10^3$ for DD and MD, and $1 \times 10^5$ for KPM. More details about the simulations can be found in ~\cite{Suppl} including Refs.~\cite{stiffness,Rahman1976,LAMMPS,Milkus2018,Taraskin1997}}
 \label{fig:DOS}
\end{figure}

\section{Viscoelastic response from nonaffine lattice dynamics}
The nonaffine lattice dynamics (NALD) approach \cite{Lemaitre2006} assumes that the deformation can be represented as a sum of two contributions: (1) the affine deformation typical for centrosymmetric materials, and (2) a nonaffine relaxation of atomic positions towards new equilibrium positions. The latter part produces a negative correction to the elastic free energy. Here we specialize to shear deformations, keeping in mind that the equations can be easily rewritten for other types of deformation. The expression for the free energy of deformation reads~\cite{Prediction2018}
\begin{equation}
F=F_\mathrm{A}-F_\mathrm{NA}=F_\mathrm{A}-\frac{1}{2}\sum_i\frac{\partial\mathbf{f}_i}{\partial\gamma} \cdot \frac{\partial \mathbf{r}_i}{\partial\gamma}\gamma^2,
\label{FreeEnergy1}
\end{equation}
where $F_\mathrm{A}$ is the affine contribution and $-F_\mathrm{NA}$ is the non-affine contribution. The latter can be expressed through derivatives of net force due to affine deformation, the derivatives of the radius-vector and the shear strain amplitude (angle) $\gamma$. Under constraint of mechanical equilibrium and small deformations ($\gamma \to 0$) Equation \ref{FreeEnergy1} can be written as~\cite{Zaccone2011}:
$F=F_\mathrm{A}-\frac{1}{2}\mathbf{\Xi}_iH^{-1}_{ij}\mathbf{\Xi}_j\gamma^2$,
where we introduce the variable $\mathbf{\Xi}_i$ for the affine force field, defined through the force acting on atom $i$, $\mathbf{f}_i=\mathbf{\Xi}_i\gamma$, while $H_{ij}$ is the Hessian of the system (a $3N\times3N$ matrix), and summation over repeated indices is implied. The assumption of small deformations means that the system resides in the vicinity of a local energy minimum (effects of anharmonicity and temperature are taken into account via tension terms and negative eigenvalues of the Hessian ~\cite{Prediction2018,ElderZacconeSirk:2019}). With dissipation at the molecular level, the equation of motion for a particle $i$ of mass $m$ can be written in damped harmonic oscillator form \cite{Lemaitre2006},
\begin{equation}
m\ddot{\mathbf{r}}_{i}+\nu\dot{\mathbf{r}}_{i}+H_{ij}\mathbf{r}_{j}=\mathbf{\Xi}_{i}\gamma,
\label{oscillator}
\end{equation}
with inertial, dissipative and harmonic force terms on the left hand side and the affine-force field on the right side. This key equation in the non-affine formalism has to be modified for the multi-component case. In the multi-component case the particles or atoms can have different masses. Hence, the equation (\ref{oscillator}) can be rewritten in the following form \cite{JohannesThesis},
\begin{eqnarray}\label{eq:damped_newton}
\mathbf{M}\ddot{\mathbf{r}}(t)+\mathbf{C}\dot{\mathbf{r}}(t)+\mathbf{H}\mathbf{r}(t)=\mathbf{f}(t),
\end{eqnarray}
where $\mathbf{M}$ is a mass matrix ($N \times N$ block matrix, where each $3\times 3$ block assigns the mass $m_i$ to the particle with label $i$), and $\mathbf{r}(t)$ represents the full configuration of the system, i.e. it is a $3N$-element vector.

In order to solve Eq.~\eqref{eq:damped_newton}, let us consider an auxiliary generalized eigenvalue problem,
\begin{eqnarray}\label{eq:genEvalProb}
\omega^2_p\mathbf{M}\gEigV_p=\mathbf{H}\gEigV_p,
\end{eqnarray}
with $\gEigV_p$ and $\omega^2_p$ being eigenvectors/eigenvalues correspondingly.
Transforming this equation by multiplying with $\mathbf{M}^{-1/2}$ from the left gives
\begin{eqnarray}
\omega^2_p\mathbf{M}^{1/2}\gEigV_p=\mathbf{M}^{-1/2}\mathbf{H}\gEigV_p           
\end{eqnarray}
After inserting the unit matrix $\mathbb{I}=\mathbf{M}^{-1/2}\mathbf{M}^{1/2}$ on the right hand side
\begin{eqnarray}
\omega^2_p {\EigV}_p={\mathbf{\MH}}{\EigV}_p,       
\end{eqnarray}
where we have defined $\mathbf{\mathcal{H}}=\mathbf{M}^{-1/2}\mathbf{H}\mathbf{M}^{-1/2}$ and $\EigV_p=\mathbf{M}^{1/2}\gEigV_p$.

Now we reformulate the generalized eigenvalue problem given in Eq.~\eqref{eq:damped_newton} in a matrix form by introducing an auxiliary matrix $\mathbf{\Phi}$ with columns being made of the eigenvectors of Eq.~\eqref{eq:genEvalProb}. This matrix is related to $\mathbf{M}$ and $\mathbf{H}$ as
\begin{equation}
\mathbf{\Phi}^T \mathbf{M} \mathbf{\Phi} =1, ~~~
\mathbf{\Phi}^T \mathbf{H} \mathbf{\Phi} =\mathbf{\Omega}^2,
\label{eq:gen_eig}
\end{equation}
where we have defined another auxiliary diagonal matrix $\mathbf{\Omega}$ of the eigenfrequencies, $\mathbf{\Omega} = \text{diag}(\omega_1,\dots,\omega_{3N})$.
In order to solve the Eq.~\eqref{eq:damped_newton}  we replace $\mathbf{r} = \mathbf{\Phi}\mathbf{q}$ and multiply the equation from the left with $\mathbf{\Phi}^T$ obtaining \cite{Veselic2011}
\begin{eqnarray}\label{eq:damped_newton2}
\mathbf{\Phi}^T\mathbf{M}\mathbf{\Phi}\ddot{\mathbf{q}}+\mathbf{\Phi}^T\mathbf{C}\mathbf{\Phi}\dot{\mathbf{q}}+\mathbf{\Phi}^T\mathbf{H}\mathbf{\Phi} \mathbf{q}=\mathbf{g}
\end{eqnarray}
with $\mathbf{g}=\mathbf{\Phi}^T\mathbf{f}$ being the transformed driving force.
The first and the third term now can be substituted from Eq.~\eqref{eq:gen_eig} and simplified as,
\begin{eqnarray}\label{eq:damped_newton3}
\ddot{\mathbf{q}}+\mathbf{\Phi}^T\mathbf{C}\mathbf{\Phi}\dot{\mathbf{q}}+\mathbf{\Omega}^2 \mathbf{q}=\mathbf{g.}
\end{eqnarray}
The second term contains the matrix product $\mathbf{\Phi}^T\mathbf{C}\mathbf{\Phi}$, which makes the general analytical solution of Eq. (\ref{eq:damped_newton3}) impossible. This can be overcome by assuming that the damping is not correlated across different eigenmodes, i.e. $\mathbf{\Phi}^T\mathbf{C}\mathbf{\Phi}$ is a diagonal matrix. The frictional drag force is proportional to the mass $\mathbf{C} \propto \mathbf{M}$, which decouples the equations (\ref{eq:damped_newton3}) of motion~\cite{Veselic2011}. 
If the friction matrix $\mathbf{C}$ has non-zero off-diagonal elements one could approximate it with a diagonal matrix and check under which conditions the off-diagonal elements are small enough. 

This allows us to use index-independent notation $\hat{\nu}$ for the friction since $(\mathbf{\Phi}^T\mathbf{C}\mathbf{\Phi})_{kk}=(\mathbf{\Phi}^T\hat{\nu}\mathbf{M}\mathbf{\Phi})_{kk}=\hat{\nu}$ for any $k$. Hence we obtain a system of decoupled equations,
\begin{eqnarray}
\ddot{q}_k+\hat{\nu}\dot{q}_k+\omega_k^2 q_k=g_k.
\end{eqnarray} 
Applying a Fourier transform maps the equation to the frequency-space
\begin{eqnarray}\label{eq:damped_solution}
\widetilde{q}_k=\dfrac{\widetilde{g}_k}{-\omega^2+i\hat{\nu}\omega+\omega_k^2},
\end{eqnarray}  
where $\widetilde{q}_k,\widetilde{g}_k$ are the corresponding Fourier transforms of $q_k$ and $g_k$.

In Ref. \cite{Lemaitre2006} the general relation between the stress response $\Delta\tilde t_\eta(\omega)$ of the system to a strain $\eta$ and the displacement fields  $\mathbf{r}$ is
\begin{eqnarray}\label{eq:damped_solution2}
\Delta \widetilde{t}_\eta(\omega)=G_\text{A} \widetilde{\eta}(\omega)-\dfrac{1}{V}\sum_{i=1}^N\mathbf{\Xi}^T_i\cdot\widetilde{\mathbf{r}}_i(\omega),
\end{eqnarray}
where the summation extends over all particles. The vectors of $3N$-dimensional affine force $\mathbf{\Xi}$ and the Fourier transform of displacement field $\widetilde{\mathbf{r}}$ are functions of the driving frequency $\omega$.

In our case the second term in Eq.~\eqref{eq:damped_solution2} can be transformed by using the definition $\mathbf{r} = \mathbf{\Phi}\mathbf{q}$ and Eq. \eqref{eq:damped_solution}, 
\begin{eqnarray}
\dfrac{1}{V}\sum_{i=1}^N\mathbf{\Xi}_i^T\cdot\widetilde{\mathbf{r}}_i=\dfrac{1}{V}\sum_{i=1}^N\mathbf{\Xi}^T_i \cdot \left( \sum _p\mathbf{\Phi}_{ip}  \widetilde{\mathbf{q}}_p\right)
\notag \\=\dfrac{1}{V}\sum_i\sum_p\dfrac{\mathbf{\Xi}_i\mathbf{\Phi}_{ip}\widetilde{\mathbf{g}}_p}{-\omega^2 + i \hat{\nu}\omega + \omega_p^2}.
\end{eqnarray}
In frequency space the generalized force vector can be written as $\widetilde{\mathbf{g}}_p=\sum_{j}\mathbf{\Phi}^T_{pj}\widetilde{\mathbf{f}}_{j}$. For small deformations one can assume that the contributions of components of driving force with different frequencies are independent. Hence it is conventional to consider the case of the driving force defined as $\mathbf{f}(t) = \mathbf{\Xi}\,\widetilde{\eta}\sin\omega t$ \cite{Lemaitre2006}. The previous expression can be modified further,
\begin{equation}
\dfrac{1}{V}\sum_{i=1}^N\mathbf{\Xi}^T_i\cdot\widetilde{\mathbf{r}}_i(\omega)
=\dfrac{1}{V} \sum_{i,j} \sum_{p}\dfrac{ \left( \mathbf{\Phi}^T_{pi} \mathbf{\Xi}_i\right)^T\cdot\left(\mathbf{\Phi}^T_{pj} \mathbf{\Xi}_j \right) }{-\omega^2 + i \hat{\nu}\omega + \omega_p^2}\widetilde{\eta}(\omega).
\end{equation}
The matrix product $\sum_i\mathbf{\Phi}^T_{pi} \mathbf{\Xi}_{i}=\mathbf{\Xi}_p$ and its transposed counterpart represent the basis transformation of the affine force field into the generalized eigenbasis. Thus,
\begin{eqnarray}
\Delta \widetilde{t}_\eta(\omega)=G_\text{A} \widetilde{\eta}(\omega)-\dfrac{1}{V} \sum_{p}\dfrac{\mathbf{\Xi}^T_p  \cdot \mathbf{\Xi}_p }{-\omega^2 + i \hat\nu\omega + \omega_p^2}\widetilde{\eta}(\omega).
\end{eqnarray}
Since $\Delta \widetilde{t}_\eta(\omega) = G^*(\omega) \widetilde{\eta}(\omega)$ in the linear regime we get the final expression for the complex viscoelastic shear modulus of a multi-component disordered system
\begin{eqnarray}\label{complexmodulus}
G^*(\omega)=G_\text{A}-\dfrac{1}{V} \sum_{p}\dfrac{\mathbf{\Xi}^T_p  \cdot \mathbf{\Xi}_p }{-\omega^2 + i \hat{\nu}\omega + \omega_p^2}.
\end{eqnarray}

In the thermodynamic limit, it can be rewritten as \cite{Prediction2018},
\begin{eqnarray}\label{complexmodulus2}
G^*(\omega)=G_\text{A}-\dfrac{3N}{V} \int_{C}\dfrac{\Gamma(\omega')\rho(\omega')}{-\omega^2 + i \hat{\nu}\omega + \omega'^2}d\omega',
\label{eq:Gamma_contin}
\end{eqnarray}
where $C$ is an integration contour which includes negative eigenvalues (imaginary frequencies), widely known as instantaneous normal modes (INMs)~\cite{Stratt1995,Keyes1997,Prediction2018,Douglas2019}, $\omega'$  denotes the eigenfrequency as continuous variable, $\rho(\omega')$ is the vibrational density of states (VDOS), and the correlator $\Gamma(\omega')$ is defined in the following (note that in the last expression the mass dependence enters through $\Gamma(\omega')$). Importantly, one can see that, in the multi-component case, the expression is very similar to the single component one ~\cite{Lemaitre2006, Prediction2018}, however, the $\Gamma(\omega')$ in Eq. (\ref{eq:Gamma_contin}) has a dimensionality difference of $M^{-1}$ with respect to $\Gamma(\omega')$ from Refs. \cite{Lemaitre2006,Prediction2018}.

The results for the comparison between the VDOS computed using the Direct Diagonalization (DD) of the Hessian method and the KPM method (that will be introduced below) for the multi-mass KG polymer can be found in the Appendix F.

\section{Simulation details}

We have used the Kremer-Grest model~\cite{Kremer1986} of a coarse-grained polymer system consisting of linear chains of 50 monomers which were equilibrated using LAMMPS~\cite{LAMMPS}. 
The polymer chain under consideration consisted of two different types of masses, where the two masses were chosen as $m_1=1$ and $m_2=3$. The geometry of the chain is such that the  masses are placed in alternating fashion, as illustrated in Fig.1a.

The polymer chains are embedded in a three-dimensional box subject to periodic boundary conditions. In the Kremer-Grest model each constituent monomer is allowed to interact via a Lennard-Jones potential
\begin{equation}
U_\text{LJ}(r) = 4\epsilon\left[\left(\frac{\sigma}{r}\right)^{12}-\left(\frac{\sigma}{r}\right)^6-\left(\left(\frac{\sigma}{r_c} \right)^{12}-\left(\frac{\sigma}{r_c}\right)^6\right)\right],
\end{equation}  
where the parameters are chosen as $\epsilon=1$, $\sigma=1$. The cutoff radius of the potential is set to $r_c=2.5$.            
In addition, in the Kremer-Grest model, the covalent along-the-chain bonds are represented by a finite extensible nonlinear elastic (FENE) potential given by~\cite{Kremer1986}
\begin{eqnarray}
U_\text{FENE}(r)=-\dfrac{Kr_0^2}{2}\ln\left[1-\left(\dfrac{r}{r_0}\right)^2\right].
\end{eqnarray}
The interaction parameters of the FENE interaction are $K=30$ and $r_0=1.5$. A Langevin thermostat was used for the molecular dynamics simulations where particles experience a viscous damping force proportional to the velocity. The corresponding damping constant $\xi$, which is related to the damping term appearing in the lattice dynamical equation of motion by $\xi = m\nu$. 
Using dimensionless LJ units in terms of the mass $M$, length $d$ and energy $\mathcal{\epsilon}$, we set $\sigma=1$ and $\epsilon=1$, which results in a fundamental unit of time given by $\tau = \sqrt{m \sigma^2/\epsilon}$. 


To apply the theory described above, one must first obtain a low-energy configuration of the solid. All of the quantities can then be extracted from this snapshot of the system and the interaction potentials. We will use the same simulation procedure as in Ref. \cite{Prediction2018}. In brief, the snapshots of the system are obtained using the LAMMPS simulation package~\cite{LAMMPS}. After a sufficient number of equilibration steps in a melted state at $T^\ast = 1$ the system is slowly quenched, maintaining zero external pressure using a Nose-Hoover barostat, below the glass transition temperature ($T=\Tsim \ll T_\text{g}\approx 0.4$). The timescale of the cooling is $\tau_c \gtrsim 10^5\tau$. Ten replica configurations were constructed, and all results are averaged over these ten structures. Each glassy configuration is used as an input for the calculation of the Hessian. The latter is then diagonalized directly for comparison with the distributions of the VDOS and $\Gamma$ correlator obtained by KPM. The viscoelastic moduli are also extracted from direct mechanical spectroscopy simulations \cite{Prediction2018,Rodney2017} and compared then with the theoretical predictions. 

 For the eigenanalysis via direct diagonalisation a system consisting of $N=5000$ particles. The analysis using KPM was done on a system with the same parameters but of significantly larger size of $N=1 \times 10^5$ particles which would be challenging for direct diagonalisation. We have checked that the results for the viscoelastic response of the system obtained with the KPM from the small and large system give similar results. Clearly, there will be slight variations in the VDOS and $\Gamma(\ww)$ due to the fact that different snapshots (samples) of the glass are considered. 

\begin{figure*}
\includegraphics[width=16cm]{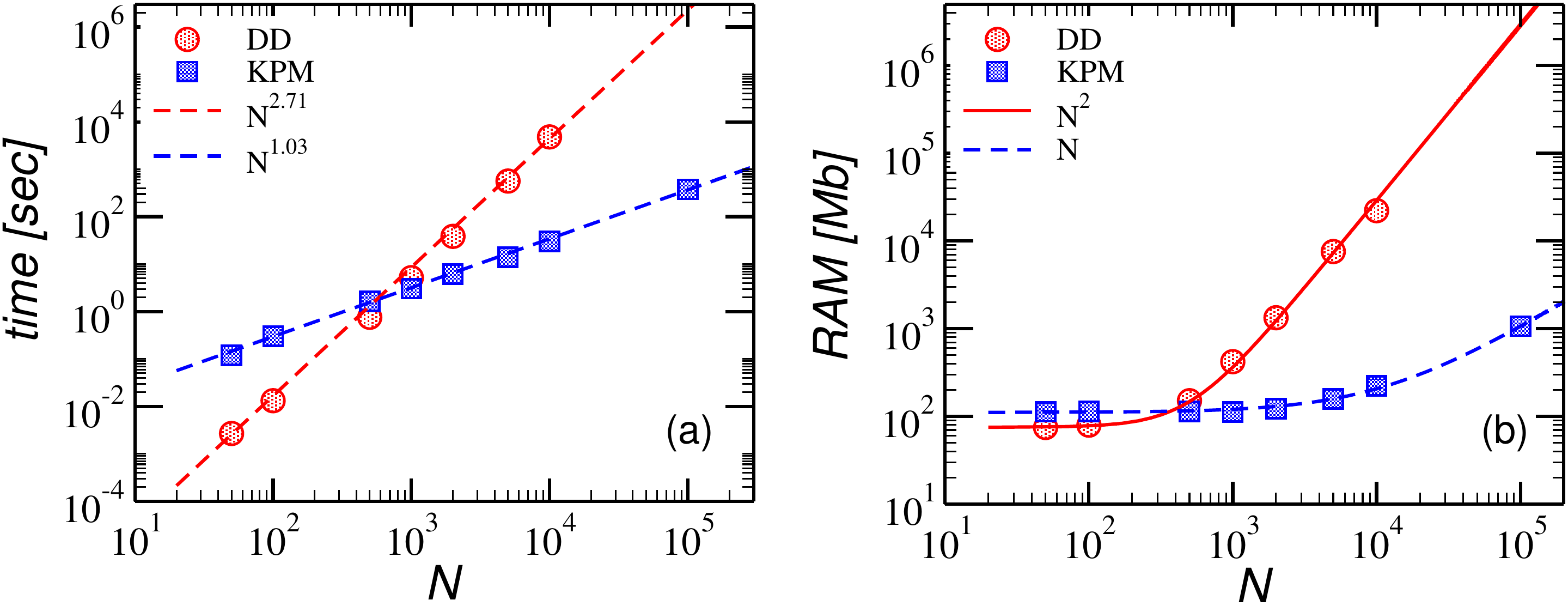}
\caption{Panel (a): Scaling of the required time with the number of particles (N) for DD and KPM (K=50, R=500, C=32). Panel (b): memory usage for DD and KPM.}
\label{fig:timeRAM}
\end{figure*}

\section{Kernel Polynomial Method to compute the eigenmodes}
The relevant quantity appearing in the expression for complex viscoelastic shear modulus is the product of the $\Gamma$-correlator and the VDOS $\rho(\omega')$. $\Gamma(\omega')$ is defined as a squared norm of the projection of the eigenvectors of the Hessian matrix onto the affine force field of the disordered particle system, i.e. $\vert\langle\mathbf{\Xi}\vert\mathbf{p}\rangle\vert^2$, where $\vert\mathbf{p}\rangle$ represents an eigenvector of the Hessian matrix~\cite{Lemaitre2006}.

We will now discuss how this quantity can be evaluated by means of the KPM methodology developed here, which allows one to scale the calculation up to much larger systems than previously possible with the direct diagonalization method. We should emphasize here that the KPM does not need the assumptions of small $\gamma$ and mechanical equilibrium, used in the theory. It does not care about the physical origin of the Hessian $H$ or of the affine force field $\Xi$.

Since the ansatz of KPM starts with the decomposition of the VDOS expressed as a sum of $\delta$-functions \cite{Weisse2006}, we will directly compute the product of $\vert\langle\mathbf{\Xi}\vert\mathbf{p}\rangle\vert^2$ and $\rho(\omega')$ with KPM. The $\Gamma$ correlator can then be computed by dividing this quantity by the VDOS.  The exact expression for the product reads~\cite{Lemaitre2006}
\begin{eqnarray}
\mathcal{J}(\omega')=&\rho(\omega')\Gamma(\omega')=\dfrac{2 \omega'}{3N}\sum_p\langle\mathbf{\Xi}\vert\mathbf{p}\rangle\langle\mathbf{p}\vert\mathbf{\Xi}\rangle\delta(\widetilde{\lambda}-\widetilde{\lambda}_p). \nonumber
\end{eqnarray}
The KPM approximation, as shown in Appendix A, gives
\begin{eqnarray}
\mathcal{J}(\omega')=\dfrac{4 \omega'}{\pi}\dfrac{2-\varepsilon}{\lambda_\text{max}-\lambda_\text{min}}\sum_{k=0}^\infty\gamma_k\mu_k\sin\big[(k+1)\arccos\widetilde{\lambda}\big], \nonumber
\end{eqnarray}
where the Chebyshev expansion coefficients $\mu_k$ are calculated as follows.

Using the notation $\vert\mathbf{u}_k\rangle=U_k({\widetilde{\mathbf{H}}})\vert\mathbf{u}_0\rangle$, 
where $U_k$ are Chebyshev polynomials of the second kind and $\widetilde{\mathbf{H}}$ is a rescaled Hessian,
as shown in detail in Appendix A, we obtain that
\begin{eqnarray}
m_k=\langle \mathbf{u}_0\vert\mathbf{\Xi}\rangle\langle\mathbf{\Xi}\vert U_k(\widetilde{\mathbf{H}})\vert\mathbf{u}_0\rangle
\end{eqnarray}
is the correct approximate Chebyshev moment which stochastically converges to $\mu_k$, i.e. $\overline{m_k}\to\mu_k$. Here, $\mathbf{u}_0$ are random vectors and the average is taken over a certain number of realizations of random vectors. These expressions are valid for a one-component system, but they have been extended here to the multi-component case. Full details of the derivation can be found in the Appendices B, C, and D. 

\section{KPM performance and comparison with direct diagonalization of the Hessian}
The two key parameters which control the convergence and the accuracy of the KPM method are: $K$, which is the degree of the Chebychev polynomial at which the sum over $k$ for $\mathcal{J}(\omega')$ is truncated, and $R$, which is the number of random vectors $\mathbf{u}_0$ realizations over which the average for $m_k$ is taken.
The full details about the analysis of convergence of the KPM procedure can be found in the Appendix E.

To be more precise, our realisation of KPM method has exactly $R \times K$ dot products of the sparse matrix and vector, and $R \times K$ dot products of two vectors. The speed of computation of these dot products depends on the details of the linear algebra libraries used and the sparsity of the Hessian matrix. However, each iteration in \textit{R-cycle} of KPM is independent, hence KPM can be parallelized over $R$ cycles. We apply this parallelization in the following calculations of the viscoelastic shear moduli $G'(\omega)$ and $G''(\omega)$ with KPM.

In Fig. 1(b)-(c), the results from MD simulations of oscillatory deformation for the polymer glass in Fig. 1(a) performed in LAMMPS are compared with the theoretical calculations using NALD with (i) $\rho(\omega')\Gamma(\omega')$ evaluated with direct diagonalization (DD) of static MD snapshots and (ii) evaluated with KPM. With DD we use $N=5000$ while with KPM we use $N=1 \times 10^5$. An excellent agreement is observed across the entire frequency range. Note that the NALD equations are fully predictive with no adjustable parameters. Even the friction parameter is taken to be identical with the friction value set in the Langevin thermostat of the MD simulations (see Section III). The successful comparison validates the derivation of multi-mass NALD above. 

In the perspective of using NALD for atomistic calculations, it is important to evaluate how the computation time and the memory usage scale with the number of particles, $N$. Panel (a) of Figure \ref{fig:timeRAM} shows the dependence of the computation time for DD and KPM methods performed on multi-mass systems with different $N$, and KPM parameters $K=50$ and $R=500$. We can see that the DD method scales almost as $N^3$, in contrast to KPM which exhibits linear dependence on $N$,
\begin{equation}
t_{KPM} \propto K \cdot R \cdot C \cdot d \propto N
\end{equation}
with $C$ being the number of cores, $d$ the density of the Hessian matrix, and one should also take into account the $N$-dependence of $d$ (see below).

Panel (b) of Figure \ref{fig:timeRAM} shows the comparison of the memory usage of DD and KPM for the same systems. As expected, the memory requirement for DD is proportional to $N^2$, whereas the KPM memory usage is proportional to $N$, or to be more exact, $d N^2 \sim N$. This is mostly due to the fact that KPM uses sparse matrices and in our case the density of the matrix $d$ is proportional to the inverse of the system size. This is the consequence of the cutoff introduced in our potential. This cutoff limits the number of interactions each atom can have to a certain number $N_{inter}$ ($\approx 68.5$ in our system). Thus, the total number of non zero Hessian elements is $9 N N_{inter}$ and the density $d=\frac{9 N N_{inter}}{N^2}=\frac{N_{inter}}{N}\propto \frac{1}{N}$. In our case, we have a simple system with no angular (bond-bending) or dihedral potentials (Fig. 1(a)). In general, the $N_{inter}$ depends not only on the cutoff  but on  the complexity of the potentials. Other interactions mean just a slightly different form of Hessian, but do not change the idea or algorithm of KPM. Actually, introduction of simple angular and dihedral potentials does not even change the Hessian density (since with our cutoff these particles already interact via LJ potential, and hence the Hessian elements are already non-zero),  thus the performance of the KPM remains unaltered.

\section{Conclusions}
In conclusion, we have developed a new KPM-based multi-mass lattice dynamics method for computing the mechanics of real materials, which scales linearly in time with $N$, as opposed to the standard direct diagonalization (DD), which scales as $N^{2.7}$. The new method is also much more efficient in terms of memory storage, with a memory consumption that scales as $\sim N$ as opposed to $\sim N^{2}$ found for DD. This methodology may prove key to solve the longstanding time-scale bridging problem of atomistic simulations, which can access only the extreme high-rate ($\sim10^{10}$Hz) response of materials, due to the shortness of time-step. With the new method proposed here it will be possible to compute the viscoelastic response of large systems ($N=10^{6}$) at atomistic or coarse-grain resolution down to deformation rates that are experimentally accessible. Furthermore, the method is general and can be applied to any solid, including perfect crystals~\cite{Cui2019} and real crystals~\cite{Rodney1999}.

\begin{acknowledgements}
A.Z. and I.K. gratefully acknowledge financial support from US Army Research Office through contract nr. W911NF-19-2-0055. Dr. Johannes Krausser is gratefully acknowledged for discussions and input during the early phase of this work.
\end{acknowledgements}

\appendix

\section{Derivation of the Kernel Polynomial Method}
We start by shortly describing the basics of the KPM algorithm starting from Ref. \cite{Weisse2006} (which summarizes the method as it was originally developed in the context of Fermionic particles). We consider a real-valued function $f(x)$ on the interval $[a,b]$. The key idea behind the kernel polynomial approximation lies in the expansion of the function $f(x)$ into a series of Chebyshev polynomials of the second kind $U_k(x)$~\cite{Weisse2006}, i.e.
\begin{eqnarray}\label{eq:cheb_series}
f(x)=\sum_{n=0}^\infty\alpha_kU_k(x). 
\end{eqnarray}
The polynomials of the second kind $U_k({x})$ are used, because they show better convergence properties than the polynomials of the first kind \cite{Weisse2006}.

Introducing the weighted scalar product on the interval $[-1,1]$, we have
\begin{eqnarray}
\langle f\vert g\rangle_\xi=\int_{-1}^1\xi(x)f(x)g(x)dx.
\end{eqnarray}  
The Chebyshev polynomials of the second kind are orthogonal with respect to the weight $w(x)=\pi\sqrt{1-x^2}$, i.e.      
\begin{eqnarray}
\langle U_k\vert U_l\rangle_w = \dfrac{\pi^2}{2}\delta_{k,l},
\end{eqnarray}
where $\delta_{k,l}$ represents the Kronecker delta. Hence, the expansion coefficients $\alpha_k$ appearing in Eq.~\eqref{eq:cheb_series} are given by
\begin{eqnarray}
\alpha_k=\dfrac{2}{\pi}\int_{-1}^1\sqrt{1-x^2}U_k(x)f(x)dx. 
\end{eqnarray}
The Chebyshev polynomials $U_k(x)$ can also be computed using the recurrence relations
\begin{eqnarray}\label{eq:cheb_recur}
U_0(x)&=1,
\notag \\ 
U_1(x)&=2x,
\notag \\
U_k(x)&=2xU_{k-1}(x)-U_{k-2}(x)
\end{eqnarray}
or, equivalently, can be defined through their trigonometric representation
\begin{eqnarray}
U_k(x)=\dfrac{\sin\big[(k+1)\arccos x\big]}{\sqrt{1-x^2}}.
\end{eqnarray}  

\section{KPM for computation of the vibrational density of states}
One of the first applications of the KPM algorithm in physics was the computation of the eigenfrequency spectrum of a generic Hessian matrix \cite{Weisse2006,Beltukov2016,JohannesThesis}. The vibrational density of states (VDOS) can be defined as
\begin{equation}
\rho(\ww)=\dfrac{1}{3N}\sum_p\delta(\ww-\wwp).
\end{equation}
Here and in the following we use $\ww$ to denote the eigenfrequency as a continuous variable, and $\omega_p$ to denote the eigenfrequency as a discrete variable.
For the KPM we have to express it as a series of Chebyshev polynomials. The function $\rho(\ww)$ is the distribution of eigenfrequencies which result from the generic eigenvalue problem $\mathbf{H} \mathbf{x}_p = \lambda_p \mathbf{x}_p$. Usually the matrix $\mathbf{H}$ represents the Hessian matrix of an interacting particle system, where the eigenvalues represent the vibrational eigenfrequencies, i.e. $\lambda_p = \wwp^2$.
Since the set of eigenvalues $\{\lambda_p\}_{p\in 1,\dots,3N}$ of the underlying $3N \times 3N$ Hessian matrix are just the squared eigenfrequencies, we can use the variable transformation $\lambda = \ww^2$ and write the DOS as
\begin{equation}
\rho(\ww)=\dfrac{2\ww}{3N}\sum_p\delta(\ww^2-\wwp^2)
=\dfrac{2 \ww}{3N}\sum_p \delta(\lambda - \lambda_p).
\end{equation}
In order to be able to apply the KPM algorithm, the support of the function $\rho(\ww)$ has to be mapped onto the interval $[-1,1]$, to allow the expansion in terms of Chebyshev polynomials. We thus need to express the VDOS in terms of a rescaled variable $\widetilde{\lambda}$, such that the original support of eigenvalues $[\lambda_\text{min},\lambda_\text{max}]$ is mapped onto $[-1, 1] \ni \widetilde{\lambda}$. This can be achieved by a linear transformation of the eigenvalue problem, which is given by~\cite{Weisse2006}
\begin{eqnarray}\label{rescaling}
\widetilde{\mathbf{H}}=\dfrac{\mathbf{H}-b}{a}\\
\widetilde{\lambda}=\dfrac{\lambda-b}{a}\\
a=\dfrac{\lambda_\text{max}-\lambda_\text{min}}{2-\varepsilon}\\
b=\dfrac{\lambda_\text{max} + \lambda_\text{min}}{2}.
\end{eqnarray}
where $\varepsilon$ is a small parameter which has the function of stabilising the convergence of the kernel polynomial method against unwanted fluctuations at the edges of the support of the eigenvalue spectrum, known as Gibbs oscillations\cite{Weisse2006}. The extremal eigenvalues $\lambda_{min}$ and $\lambda_{max}$ can easily be found by standard Lanczos or Arnoldi algorithms. Using the above transformation we can express the VDOS as
\begin{eqnarray}\label{eq:DOS_KPM1}
\rho(\ww)=\dfrac{2 \ww}{3N}\dfrac{2- \varepsilon}{\lambda_\text{max} - \lambda_\text{min}}\sum_j \delta(\widetilde{\lambda} - \widetilde{\lambda}_j).
\end{eqnarray}  
We now just have to expand the $\delta$-function appearing in Eq.~\eqref{eq:DOS_KPM1} in terms of the Chebyshev polynomials $U_k(\widetilde{\lambda})$.
Making use of the relation $\int f(y) \delta(x-y)dy =f(x)$ we can express the $\delta$-function as~\cite{Beltukov2016}
\begin{eqnarray}
\delta(\widetilde{\lambda} - \widetilde{\lambda}_p)=\dfrac{2}{\pi}\sqrt{1-\widetilde{\lambda}^2}\sum_{k=0}^\infty U_k(\widetilde{\lambda}) U_k(\widetilde{\lambda}_p).
\end{eqnarray}
Using the trigonometric definitions of the Chebyshev polynomials one can write the series expansion~\cite{Beltukov2016},
\begin{equation}\label{eq:DOSKPM_approx1}
\rho(\ww)=\dfrac{4 \ww}{\pi}\dfrac{2- \varepsilon}{\lambda_\text{max} - \lambda_\text{min}}\sum_{k=0}^\infty\mu_k
\sin\big[(k+1)\arccos\widetilde{\lambda}\big],       
\end{equation}
where we have introduced the Chebyshev moments defined by
\begin{eqnarray}\label{eq:Cheb_moment}
\mu_k=\dfrac{1}{3N}\sum_{j=1}^{3N}U_k(\widetilde{\lambda}_j).
\end{eqnarray}  
The approximation then essentially consists of truncating the infinite series at a finite order,
\begin{eqnarray}
\delta(\widetilde{\lambda}-\widetilde{\lambda}_j)\approx\dfrac{2}{\pi}\sqrt{1-\widetilde{\lambda}^2}\sum_{k=0}^K\gamma_kU_k(\widetilde{\lambda}_j) U_k(\widetilde{\lambda}).
\end{eqnarray}

At this point the damping factor $\gamma_k$ has to be introduced to counteract thed Gibbs oscillations. The damping induced by $\gamma_k$ effectively truncates the series expansion gradually to avoid the oscillatory fluctuations which would appear if the sum were truncated abruptly~\cite{Beltukov2016,Weisse2006}. By substituting this truncated series into the expression for the VDOS in Eq.~\eqref{eq:DOS_KPM1} we obtain the approximate VDOS as
\begin{equation}\label{eq:DOSKPM_approx}
\rho(\ww)=\dfrac{4 \ww}{\pi}\dfrac{2- \varepsilon}{\lambda_\text{max} - \lambda_\text{min}}\sum_{k=0}^K\gamma_k\mu_k\sin\big[(k+1)\arccos\widetilde{\lambda}\big]   
\end{equation}
where $K$ denotes the degree of the approximation which basically sets the resolution of the algorithm for approximating the $\delta$-peaks which constitute the VDOS. 

The moments $\mu_k$ can be found from a modification of Eq.~\eqref{eq:Cheb_moment},
\begin{eqnarray}\label{eq:kpm_pull_evec}
\mu_k=\dfrac{1}{3N}\sum_{p=1}^{3N}\langle\mathbf{p}\vert U_k(\widetilde{\mathbf{H}})\vert\mathbf{p}\rangle,
\end{eqnarray}
where $\vert\mathbf{p}\rangle$ represent normalised eigenvectors of the rescaled Hessian matrix $\widetilde{\mathbf{H}}$. The central point of the KPM is that the above trace can be approximated stochastically very accurately if the matrix $\widetilde{\mathbf{H}}$ becomes very large~\cite{Weisse2006}. Thus, instead of evaluating the trace over the full set of all eigenvectors, we initialise a number of normalised Gaussian random vectors $\vert\mathbf{u}_0\rangle$, which we want to use for the evaluation of the above trace. As an example, let us first expand one realisation of the Gaussian random vector in terms of the eigenvectors of the matrix $\widetilde{\mathbf{H}}$, i.e.
\begin{eqnarray}\label{eq:gaussian_av}
\vert\mathbf{u}_0\rangle=\sum_p\vert\mathbf{p}\rangle\langle\mathbf{p}\vert\mathbf{u}_0\rangle=\sum_p\alpha_p\vert\mathbf{p}\rangle.
\end{eqnarray}
Hence, using this expansion we obtain the matrix elements   
\begin{eqnarray}
\langle\mathbf{u}_0\vert U_k(\widetilde{\mathbf{H}})\vert\mathbf{u}_0\rangle=\sum_{p=1}^{3N}\vert\alpha_p\vert^2U_k(\widetilde{\lambda}_p),
\end{eqnarray}
which holds due to the orthonormality of the eigenvectors. The components of the random vector $\vert\mathbf{u}_0\rangle$ in an arbitrary basis, i.e. both the components $u_{0,i}$ and $\alpha_p$, are independently and identically distributed. They have zero expectation value and unit variance, i.e. $\overline{\alpha_p}=0$ and $\overline{\alpha_p^\ast \alpha_q} =\delta_{p,q}/(3N)$, where $\langle\dots\rangle$ denotes the expectation value with respect to the Gaussian probability distribution. Therefore, taking the expectation value of Eq.~\eqref{eq:gaussian_av}, we obtain
\begin{eqnarray}
\overline{\langle\mathbf{u}_0\vert U_k(\widetilde{\mathbf{H}})\vert\mathbf{u}_0\rangle}
=&\sum_{p=1}^{3N}\overline{\vert\alpha_p\vert^2} U_k(\widetilde{\lambda}_p)
\notag\\=&\dfrac{1}{3N}\sum_{p=1}^{3N}U_k(\widetilde{\lambda}_p) = \mu_k  
\label{eq:KPM_R_cycle}      
\end{eqnarray}
since the random vectors $\mathbf{u}_0$ are normalised to one, i.e. $\overline{\vert\alpha_p\vert^2} = 1/(3N)$~\cite{Beltukov2016}. Hence we can stochastically approximate the Chebyshev moments as $\mu_k \approx \overline{\langle\mathbf{u}_0\vert U_k(\widetilde{\mathbf{H}})\vert\mathbf{u}_0\rangle}$.

Upon setting $\vert\mathbf{u}_k\rangle=U_k({\widetilde{\mathbf{H}}})\vert\mathbf{u}_0\rangle$ and $m_k=\langle\mathbf{u}_0\vert\mathbf{u}_k\rangle$ we see that, after averaging over many realisations of the random vector $\vert\mathbf{u}_0\rangle$, $m_k$ will converge to $\mu_k$, i.e. $\overline{m}_k\to \mu_k$~\cite{Beltukov2016}. The relative error of the stochastic approximation of the trace is of the order $\mathcal{O}(\sqrt{RN})$~\cite{Weisse2006}, where $R$ is the number of random vectors drawn from the Gaussian ensemble. 
Therefore, starting from $\vert\mathbf{u}_0\rangle$, we can subsequently compute the Chebyshev moments $\mu_k$ by applying the recurrence relation defining the Chebyshev polynomials. In the first iteration $\vert\mathbf{u}_1\rangle$ is obtained by using Eq.~\eqref{eq:cheb_recur}
\begin{eqnarray}
\vert\mathbf{u}_1\rangle =2\widetilde{\mathbf{H}}\vert\mathbf{u}_0\rangle
\end{eqnarray}
and by applied the procedure recurrently
\begin{eqnarray}
\vert\mathbf{u}_k\rangle=2\widetilde{\mathbf{H}}\vert\mathbf{u}_{k-1}\rangle-\vert\mathbf{u}_{k-2}\rangle.
\end{eqnarray}

\section{KPM algorithm for the nonaffine correlator $\Gamma(\ww)$}

The relevant quantity appearing in the expression for complex viscoelastic shear modulus Eq. (\ref{complexmodulus2}) is the product of the $\Gamma$-correlator and the VDOS $\rho(\ww)$. The function $\Gamma(\ww)$ is defined as a squared norm of the projection of the eigenvectors of the Hessian matrix onto the affine force field of the disordered particle system, i.e. $\vert\langle\mathbf{\Xi}\vert\mathbf{p}\rangle\vert^2$, where again $\vert\mathbf{p}\rangle$ represents an eigenvector of the Hessian matrix.

Since the ansatz of KPM starts with the decomposition of the $\delta$-function, we will directly compute the product of $\vert\langle\mathbf{\Xi}\vert\mathbf{p}\rangle\vert^2$ and $\rho(\ww)$ with KPM. The $\Gamma(\ww)$ correlator can then be computed by dividing this quantity by the VDOS. The exact expression for the product reads
\begin{equation}
\mathcal{J}(\ww)=\rho(\ww)\Gamma(\ww)=\dfrac{2 \ww}{3N}\sum_p\langle\mathbf{\Xi}\vert\mathbf{p}\rangle\langle\mathbf{p}\vert\mathbf{\Xi}\rangle\delta(\widetilde{\lambda}-\widetilde{\lambda}_p).
\end{equation}
The KPM approximation naturally looks similar to (\ref{eq:DOSKPM_approx1}),
\begin{equation}
\mathcal{J}(\ww)=\dfrac{4 \ww}{\pi}\dfrac{2-\varepsilon}{\lambda_\text{max}-\lambda_\text{min}}\sum_{k=0}^\infty\gamma_k\mu_k\sin\big[(k+1)\arccos\widetilde{\lambda}\big],
\end{equation}
where the expansion coefficients $\mu_k$ take the form 
\begin{eqnarray}\label{eq_cheb_pol}
\mu_k=\dfrac{1}{3N}\sum_p\langle\mathbf{\Xi}\vert\mathbf{p}\rangle\langle\mathbf{p}\vert\mathbf{\Xi}\rangle U_k(\widetilde{\lambda}_p).
\end{eqnarray}
One can pull the Chebyshev polynomial $U_k(\widetilde{\lambda}_p)$ into the scalar product above in order to make use of the relation $U_k(\widetilde{\lambda}_p)\vert\mathbf{p}\rangle= U_k(\widetilde{\mathbf{H}})\vert\mathbf{p}\rangle$ (cf. Eq.~\eqref{eq:kpm_pull_evec}):
\begin{eqnarray}
\mu_k=\dfrac{1}{3N}\sum_p\langle\mathbf{p}\vert\mathbf{\Xi}\rangle\langle\mathbf{\Xi}\vert U_k(\widetilde{\mathbf{H}})\vert\mathbf{p}\rangle.   
\end{eqnarray}
In this form we obtain a trace and can deal with it by using the stochastic approximation introduced earlier. We expand a random vector $\vert\mathbf{u}_0\rangle$ with respect to eigenvectors of the transformed Hessian $\widetilde{\mathbf{H}}$ and write the statistical average of the trace as
\begin{equation}
\overline{\langle \mathbf{u}_0 \vert\mathbf{\Xi}\rangle\langle\mathbf{\Xi}\vert U_k(\widetilde{\mathbf{H}})\vert\mathbf{u}_0\rangle}=\overline{\sum_{p,q}\alpha^\ast_p \alpha_q\langle\mathbf{p}\vert\mathbf{\Xi}\rangle\langle\mathbf{\Xi}\vert U_k(\widetilde{\mathbf{H}})\vert\mathbf{q}\rangle},
\end{equation}  
where the asterisk denotes the complex conjugation. Since the components of the random vector fulfill $\overline{\alpha_p^\ast\alpha_q}=\delta_{p,q}/(3N)$, the above equation is reduced to
\begin{equation}
\overline{\langle \mathbf{u}_0 \vert\mathbf{\Xi}\rangle\langle\mathbf{\Xi}\vert U_k(\widetilde{\mathbf{H}})\vert{\mathbf{u}_0}\rangle}
=\dfrac{1}{3N}\sum_{p} \langle \mathbf{p} \vert\mathbf{\Xi}\rangle\langle\mathbf{\Xi}\vert U_k(\widetilde{\mathbf{H}})\vert\mathbf{p}\rangle=\mu_k 
\end{equation}
which takes us back to Eq.~\eqref{eq_cheb_pol} and is therefore the desired result. 
Using the same notation as for the KPM discussion of the density of states, i.e. $\vert\mathbf{u}_k\rangle=U_k({\widetilde{\mathbf{H}}})\vert\mathbf{u}_0\rangle$, we conclude that the expression
\begin{eqnarray}
m_k=\langle \mathbf{u}_0\vert\mathbf{\Xi}\rangle\langle\mathbf{\Xi}\vert U_k(\widetilde{\mathbf{H}})\vert\mathbf{u}_0\rangle.
\end{eqnarray}
is the correct approximate Chebyshev moment which stochastically converges to $\mu_k$, i.e. $\overline{m_k}\to\mu_k$. 

\section{Kernel Polynomial Method for multi-atom systems}
Various types of particles contribute differently towards the viscoelastic properties of a solid. Hence it is useful to identify the contributions of different mass species. In this subsection we define eigenvector weight functions of different mass types and partial densities of states (pDOS).

The eigenvalue distribution can be written as a sum of delta-functions, $\rho(\lambda)=1/(3N)\sum_p\delta(\lambda-\lambda_p)$. Assuming that $\vert\mathbf{p}\rangle$ is a complete set of orthonormal eigenvectors, i.e. $\langle\mathbf{p_i}\vert\mathbf{p_j}\rangle = \delta(\mathbf p_i-\mathbf p_j)$, we can express the eigenvalue distribution as
\begin{equation}
\rho(\lambda)=\dfrac{1}{3N}\sum_p\langle \mathbf{p}\vert\mathbf{p}\rangle\delta(\lambda-\lambda_p)=\dfrac{1}{3N}\sum_{p,i}\langle \mathbf{p}\vert\mathbf{i}\rangle\langle\mathbf{i}\vert\mathbf{p}\rangle\delta(\lambda-\lambda_p),
\end{equation}
where, in the last equality, we have projected the eigenvectors of the particle basis using the projection operator $\mathbb{I}=\vert\mathbf{i}\rangle\langle\mathbf{i}\vert$. Thus, $\langle\mathbf{p}\vert\mathbf{i}\rangle\langle\mathbf{i}\vert\mathbf{p}\rangle = \vert\mathbf{p_i}^2\vert$, the projection of the eigenvector onto the particle coordinate $\mathbf{i}$ provides the proportionality factor of the contribution of the vibrational motion of the $i^{\text{th}}$ degree of freedom to the full vibrational density of states. To define the generalized eigenvector weight function correctly, let us first introduce an index set $\mathcal{M}_n$ for each mass type. The set $\mathcal{M}_n$ denotes the set of labels of the particles with mass type $n$. Expressing the norm of a generalized eigenvector $\gEigV_p$ in terms of the eigenvectors $\EigV_p$, we have
\begin{eqnarray}\label{eq:genEvec}
\langle\gEigV_p\vert \gEigV_p \rangle &=& \sum_{i}\langle \gEigV_p \vert\mathbf{i}\rangle\langle\mathbf{i}\vert \gEigV_p \rangle=
\sum_{i}\langle \EigV_p \vert \mathbf{M}^{-1/2} \vert\mathbf{i}\rangle\langle\mathbf{i}\vert  \mathbf{M}^{-1/2} \vert\EigV_p \rangle
\notag\\&=&\sum_i \dfrac{1}{m_i}\vert\langle\EigV_p\vert\mathbf{i}\rangle\vert^2,
\end{eqnarray}
where $\langle\EigV_p\vert \mathbf{i}\rangle = \EigV_{p,i}$ represents the $i^\text{th}$ component of the eigenvector $\EigV_p$.    
Hence, in order to define the correct weight function in terms of the generalized eigenvectors, we need to normalise the contributions $\langle\gEigV_p\vert\mathbf{i}\rangle\langle\mathbf{i}\vert \gEigV_p\rangle$ with $\langle\gEigV_p\vert \gEigV_p \rangle$. If one uses projections on the vector components which belong to the different mass species given by the index set $\mathcal{M}_n$, $\sum_n\mathbf{P}_n=\mathbb{I}$, the  Eq.~\eqref{eq:genEvec} can be written as
\begin{eqnarray}
\sum_n\langle\gEigV_p \vert \mathbf{P}_n\vert  \gEigV_p\rangle=\sum_n \dfrac{1}{m_n}\langle\EigV_p \vert \mathbf{P}_n\vert  \EigV_p\rangle.
\end{eqnarray}
Thus, for instance, the correct weight function of mass species 1 reads
\begin{eqnarray}\label{eq:weight_func_def}
\chi_1(\ww)=\dfrac{\langle\gEigV_p \vert \mathbf{P}_1\vert  \gEigV_p\rangle}{\langle\gEigV_p\vert\gEigV_p\rangle}
=\dfrac{\frac{1}{m_1} \langle\EigV_p \vert \mathbf{P}_1\vert  \EigV_p\rangle}{\sum_n \dfrac{1}{m_n}\langle\EigV_p \vert \mathbf{P}_n\vert  \EigV_p\rangle}.
\end{eqnarray}
The generalized frequency-dependent weight functions sum up to unity, i.e.,
\begin{eqnarray}\label{eq:weights_def}
\sum_n\sum_{i \in \mathcal{M}_n} \dfrac{\|\gEigV_{p,i}(\wwp)\|^2}{\|\gEigV_{p}(\wwp)\|^2}
=\sum_n\chi_{n}(\ww)=1.
\end{eqnarray}
This gives a method for splitting the full density of states (VDOS) into different mass contributions or partial density of states (pDOS)
\begin{equation}\label{eq:weightDOS}
\rho_n(\ww)=\chi_n(\ww)\rho(\ww)
=\dfrac{1}{3N}\sum_{p=1}^{3N}\chi_n(\ww)\delta(\ww-\wwp).     
\end{equation}


In order to compute $\chi_n(\ww)$ with KPM, we have to modify the scheme used in the single-mass density of states.

In our context, the starting point for the KPM always involves summations over $\delta$-peaks. In the case of the quantities $\chi_n(\ww )$ and $\Gamma(\ww)$, the sum contains an additional weighting factor depending on eigenvectors to expand and approximate the desired function. The weight has the numerator $\langle\gEigV_p \vert\mathbf{P}_n\vert\gEigV_p\rangle$ and the denominator $\langle\gEigV_p\vert \gEigV_p \rangle$.

To implement this idea, we start with the numerator in Eq.~\eqref{eq:weight_func_def} and define an auxiliary function  $\widetilde{\chi}_n(\ww)$ \cite{JohannesThesis} 
\begin{eqnarray}
\widetilde{\chi}_n(\ww)=\dfrac{1}{3N} \sum_{p=1}^{3N}\sum_{i\in\mathcal{M}_n}\langle{\gEigV_p}\vert\mathbf{i}\rangle\langle\mathbf{i}\vert\gEigV_p\rangle\delta(\ww-\wwp),
\end{eqnarray}
which is amenable to the expansion in terms of Chebyshev polynomials. Performing the analogous steps as for the single-mass VDOS above, we now map the support of the eigenvalue spectrum of the generalized eigenvalue problem onto the interval $[-1+\varepsilon/2,1-\varepsilon/2]$ and continue with  
\begin{equation}
\widetilde{\chi}_n(\ww)=\dfrac{4\ww}{3\pi}\dfrac{2-\varepsilon}{\lambda_\text{max}-\lambda_\text{min}}\sum_{k=0}^\infty\mu_k\sin\big[(k+1)\arccos\widetilde{\lambda}\big],     
\end{equation}
where the corresponding Chebyshev moments are now
\begin{eqnarray}\label{eq:cheb_pol_weight}
\mu_k = \dfrac{1}{3N} \sum_p \sum_{i \in \mathcal{M}_n}\langle{\gEigV_p}  \vert\mathbf{i}\rangle\langle\mathbf{i}\vert\gEigV_p \rangle
U_k(\widetilde{\lambda}_p).
\end{eqnarray}
To pull the polynomials $U_k(\widetilde{\lambda}_p)$ inside the scalar product we now have to be careful since we are dealing with generalized eigenvectors. Only the eigenvector components of the mass species $n$ are projected out. Hence, it is possible to write the term $\langle{\gEigV_p} \vert\mathbf{i}\rangle\langle\mathbf{i}\vert\gEigV_p \rangle$ appearing in the above summation as $ \frac{1}{m_n}\langle{\EigV_p} \vert\mathbf{i}\rangle\langle\mathbf{i}\vert\EigV_p \rangle$, because the vectors $\mathbf{i}$ are eigenvectors of the mass matrix $\mathbf{M}$ with eigenvalue $m_n$. This allows us to use the relation $U_k(\widetilde{\lambda}_p) \vert\EigV_p\rangle= U_k(\MHKPM)\vert\EigV_p\rangle$ and obtain  
\begin{eqnarray}
\mu_k = \dfrac{1}{3N} \sum_p \sum_{i \in \mathcal{M}_n}\langle{\gEigV_p}\vert\mathbf{i}\rangle\langle\mathbf{i}\vert U_k(\MHKPM)\vert \gEigV_p \rangle 
\end{eqnarray}
by reabsorbing the factor $m_n$ into the generalized eigenvector, where $\MHKPM$ is a rescaled Hessian $\MH$ as given in Eq. \ref{rescaling}. Subsequently, making use of the stochastic evaluation of this trace with the Gaussian random vectors $\randVect_0$, the Chebyshev moments in Eq.~\eqref{eq:cheb_pol_weight} are approximated by averaging the quantity
\begin{eqnarray}
m_k=\sum_{i\in\mathcal{M}_n}\langle{\randVect_0} \vert\mathbf{i}\rangle\langle\mathbf{i}\vert\randVect_k \rangle=\langle\randVect_0\vert \mathbf{P}_n\vert\randVect_k\rangle,
\end{eqnarray}
where $\mathbf{P}_n$ represents the projector of the particle species $n$.         
The approximate Chebyshev moments converge to the actual Chebyshev moments $\mu_k$, i.e. $\overline{m}_k \to \mu_k$. Due to the fact that random vectors $\randVect_k$ are supposed to represent the generalized eigenvector of the multi-component system, it would be incorrect to use normalised Gaussian random vectors as before. 

To achieve the correct stochastic approximation of the Chebyshev moments $\mu_k$, we multiply a normalised random vector $\mathbf{\xi}_0$ by the inverse square root of the mass matrix $\mathbf{M}$. As a result, the initial random seed of the KPM algorithm in this case is the random vector $\randVect_0= \mathbf{M}^{-1/2} \mathbf{\xi}_0$.
The same reasoning is applicable to the denominator of Eq.~\eqref{eq:weight_func_def}. The first step is an auxiliary function given by
\begin{eqnarray}
\chi_\text{norm}(\ww)=\dfrac{1}{3N} \sum_{p=1}^{3N}\langle{\gEigV_p} \vert\gEigV_p \rangle \delta(\ww-\wwp),
\end{eqnarray}
where we use the subscript to signal that this is the KPM approximation function for the normalisation factor of the weight function $\chi_n(\ww)$. Going through the same steps as for $\widetilde{\chi}_n(\ww)$, the final result in terms of the associated approximate Chebyshev moment is
\begin{eqnarray}
m_k=\langle{\randVect_0}\vert\randVect_k\rangle,
\end{eqnarray}
which converges to the true Chebyshev moments appearing in the expansion of $\chi_\mathrm{norm}(\ww)$
\begin{eqnarray}
\mu_k = \dfrac{1}{3N} \sum_p \langle\gEigV_p\vert U_k(\MHKPM)\vert\gEigV_p\rangle
\end{eqnarray}
in the statistical average as $\overline{m_k}\to \mu_k$.
Having set up the KPM approximation for these two components we subsequently obtain the weight functions $\chi_n(\ww)$ as the ratio of the two converged auxiliary functions, i.e.
\begin{eqnarray}
\chi_n(\ww)=\dfrac{\widetilde{\chi}_n(\ww)}{\chi_\text{norm}(\ww)}
\end{eqnarray}
which is defined on the support of the eigenvalue spectrum of the Hessian matrix with the condition that $\chi_\text{norm}(\ww)\neq0$.

\section{Convergence properties of the Kernel Polynomial Method}
\label{sec:kpmConvergence}

\begin{figure}[!t]
\includegraphics[width=8cm]{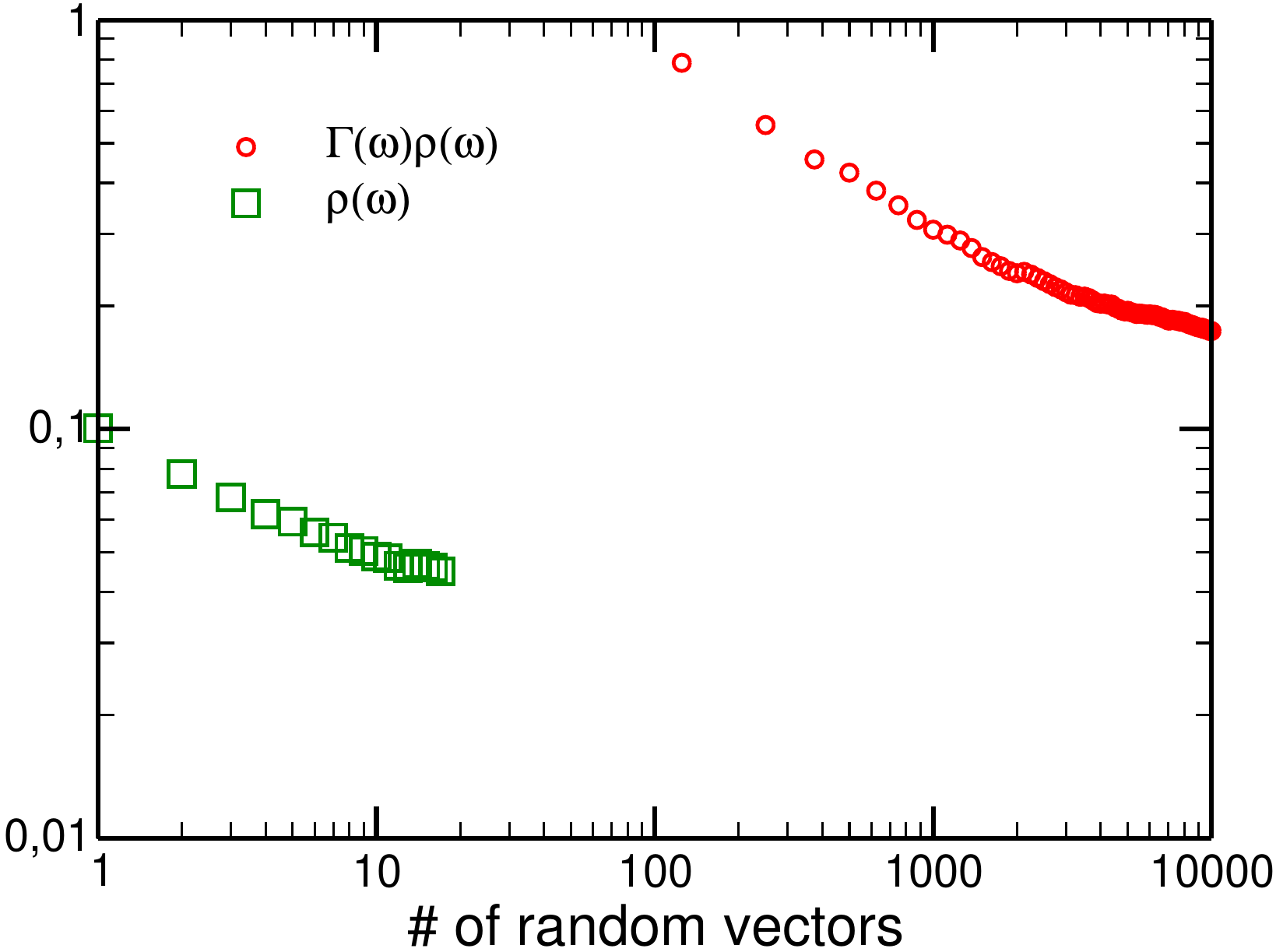}
\caption{Convergence of the KPM with the number of random vector, the difference with the reference  5k system is plottted. $K=1000$ for both DOS and $\Gamma(\ww)\rho(\ww)$.}
\label{fig:convergence}
\end{figure}

We can estimate the convergence of the KPM by the scalar difference value $\frac{\sum_i ((a_i)^2 - (a_i^{ref})^2)^{\frac{1}{2}}}{\sum_i ((a_i^{ref})^2)^{\frac{1}{2}} }$, with $a_i$ denoting is the values of $\rho(\ww)$ or $\Gamma(\ww)\rho(\ww)$ calculated with KPM, while $a^{ref}_i$ denotes the reference values of the exact solution. Using the 5k system as the reference KPM, Figure \ref{fig:convergence} shows the convergence of the KPM with the number of random vectors.
We note that $\mathcal{J}(\ww)$ has much slower convergence  than $\rho(\ww)$. Hence, a significantly larger number $R$ of sample random vectors has to be drawn in order to achieve a good approximation. The cause for this difficulty stems from the fact that the random vector used in the approximation of the Chebyshev moments $\mu_k$ is projected on the affine force field vector $\mathbf{\Xi}$, which itself is an inherently random quantity due to the structural disorder of the polymer configuration.

As a consequence, larger fluctuations occur which need more iterations to be smoothed out. To achieve a good approximation using the KPM for the VDOS, usually between 10-100 averaging iterations are required. In the case of the non-affine correlator $\Gamma(\ww)$ estimation, however, between $10^3$ to $10^4$ iterations are needed to converge the algorithm to a reasonable degree, depending also on the desired resolution.

\begin{figure*}
\centering
\begin{minipage}{2\columnwidth}\footnotesize
\centering

\subfloat[Full VDOS\label{fig:DOS_full}]{\includegraphics[width=0.45\columnwidth]{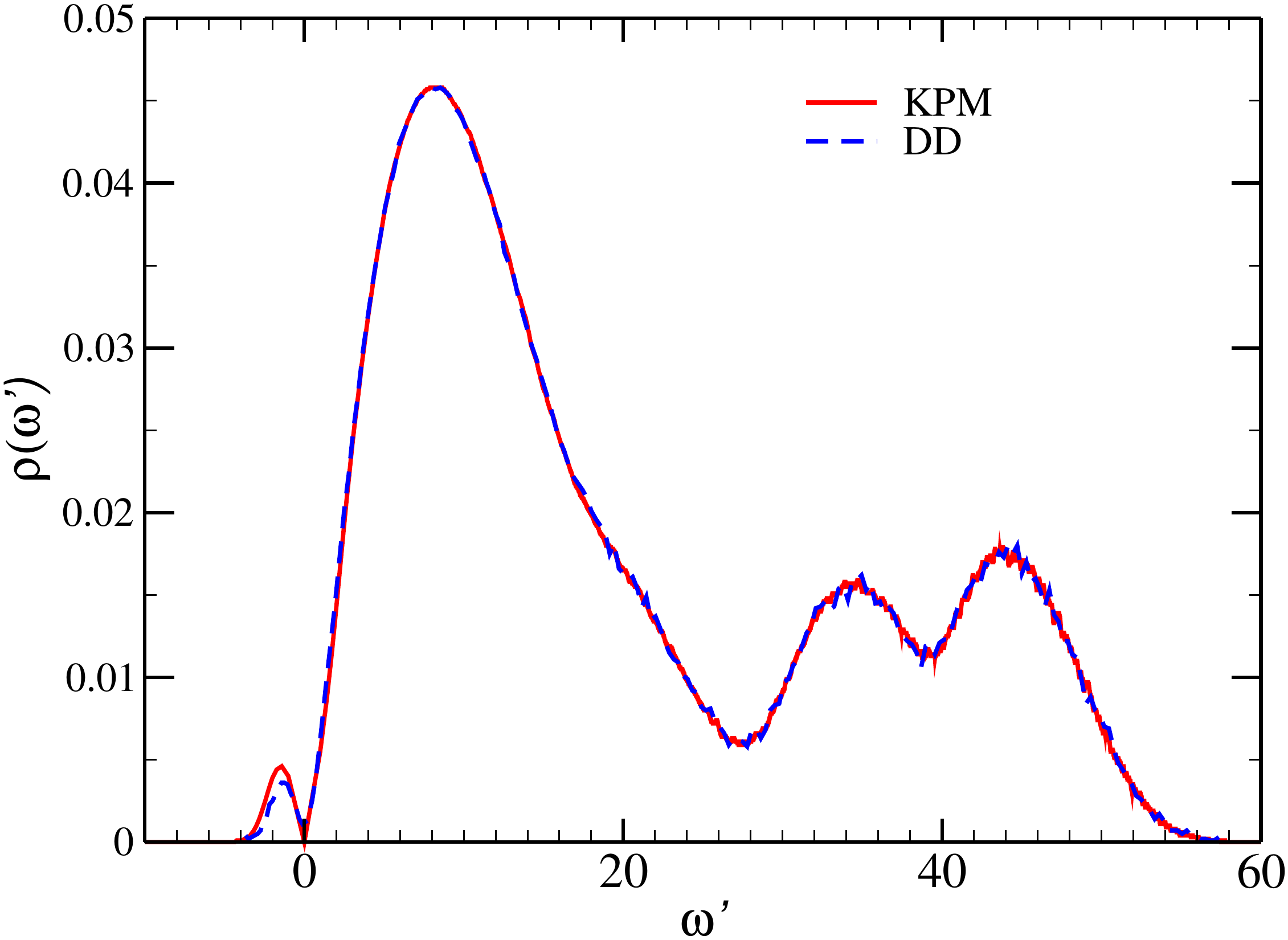}}%
\qquad
\subfloat[Partial VDOS (pDOS) for $m_1=1$\label{fig:pDOS_m1}]{\includegraphics[width=0.45\columnwidth]{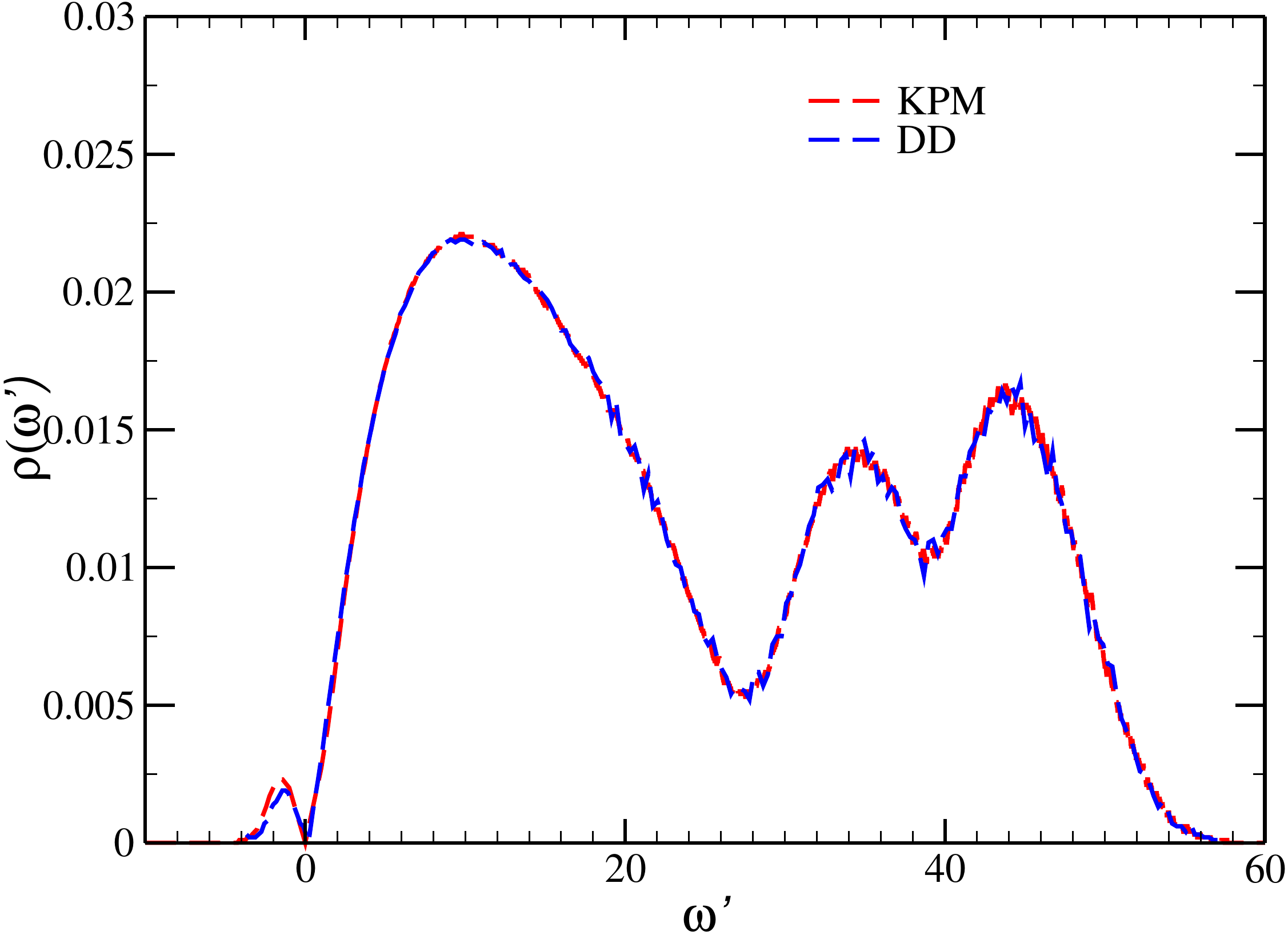}}
\qquad
\subfloat[Partial VDOS (pDOS) for $m_2=3$\label{fig:pDOS_m2}]{\includegraphics[width=0.45\columnwidth]{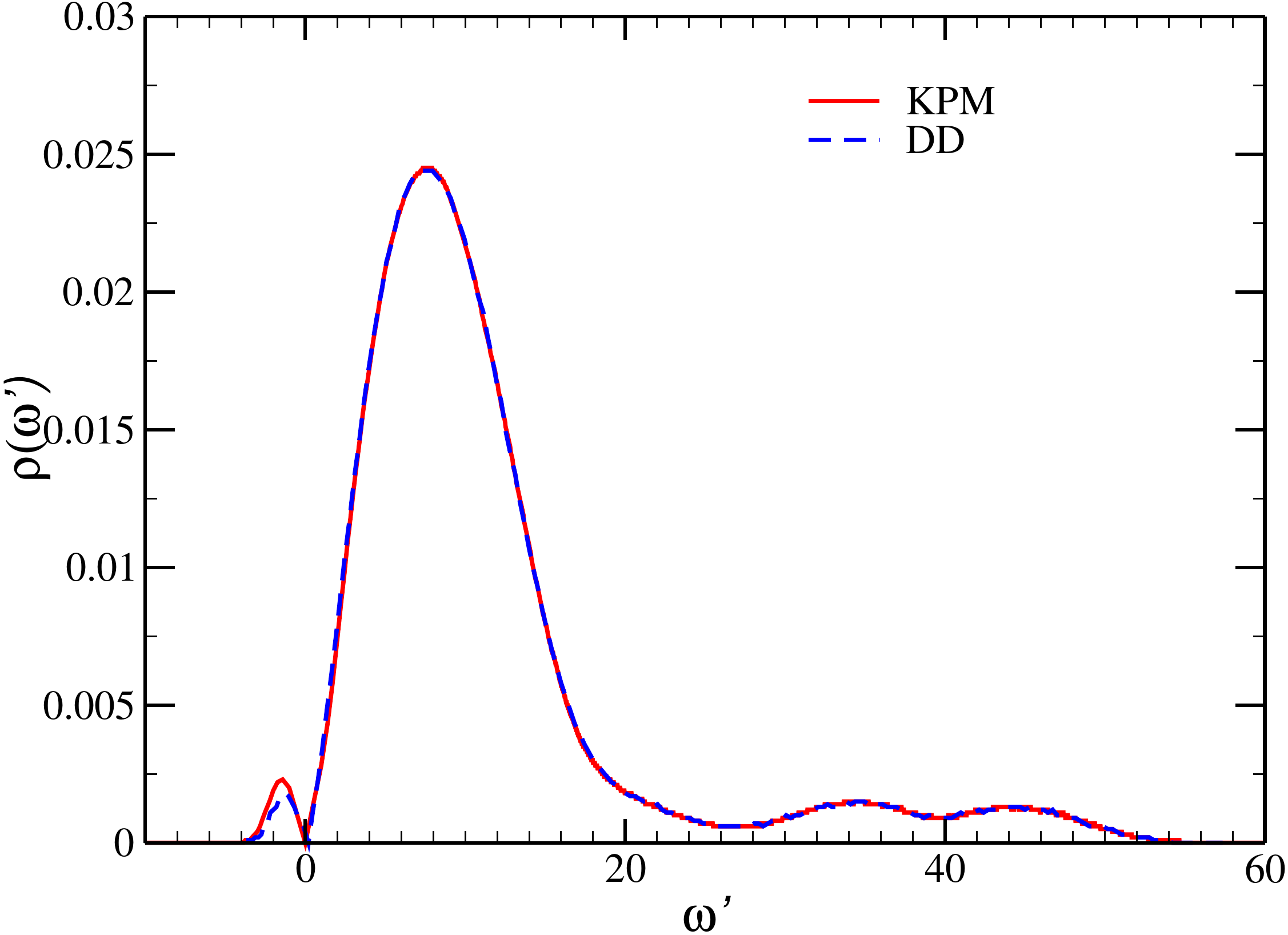}}%
\qquad
\subfloat[Full KPM VDOS with the two pDOS contributions\label{fig:3mass_fullDOS}]{\includegraphics[width=0.45\columnwidth]{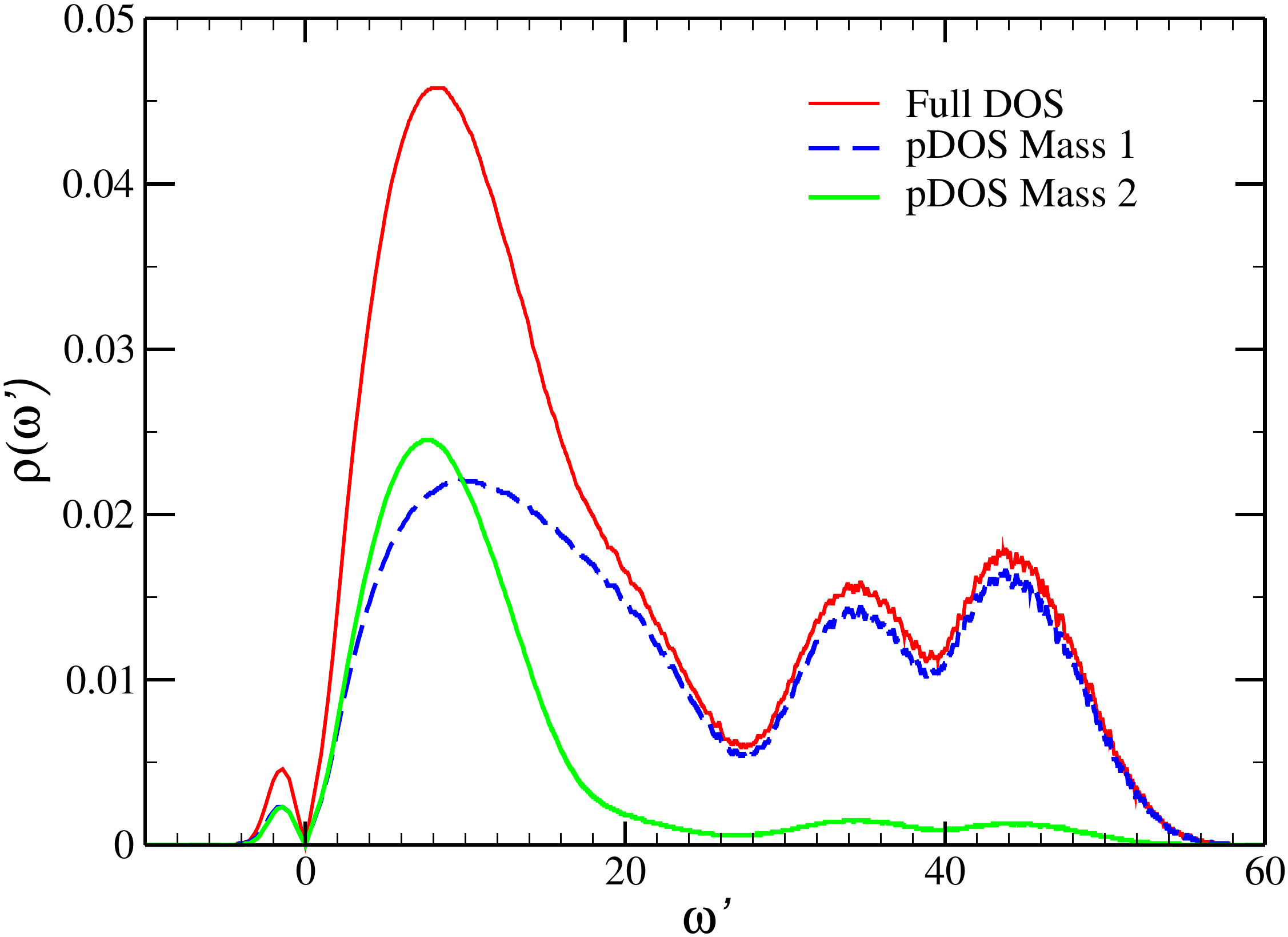}}
 \end{minipage}
 \caption{VDOS of the $T=\Tsim$ two-mass polymer glass. KPM performed on the system of $N=100000$ particles whereas DD system has only $N=5000$ particles, with $50$ monomers per chain in both cases. All results are averaged over 10 different configurations of the quenched disorder.}
 \label{fig:DOS}
\end{figure*}

\section{Results for the vibrational density of states}
\begin{figure}[t]
\includegraphics[width=8cm]{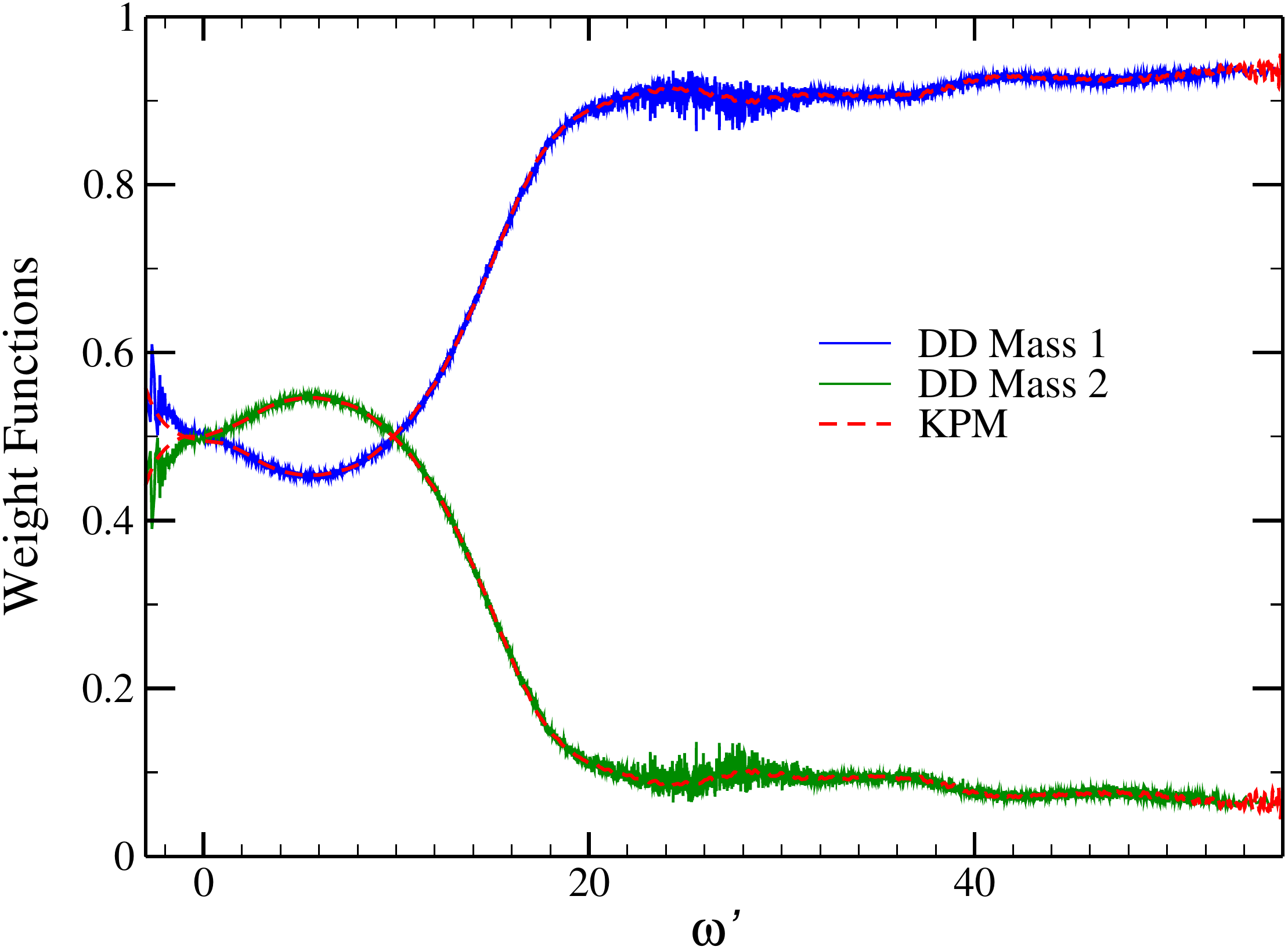}
\caption{The weight function of the two mass contributions for $T=\Tsim$ as defined in Eq.~\eqref{eq:weights_def}. The coloured points show the data obtained from direct diagonalisation (DD) for the two weight functions $\chi_n(\ww)$, $n=1,2$. The red dashed lines show the approximation to the weight function using the KPM.}
\label{fig:weightF_plot}
\end{figure}

The weight functions $\chi_n(\ww)$ are plotted in Fig.~\ref{fig:weightF_plot}. For a given mass species they represent the contribution from the species to the full eigenvector of the system at a given eigenfrequency. First, we notice that the KPM is capable of producing a very accurate approximation for $\chi_n(\ww)$. It should be noted however that the polynomial degree necessary for a good match with the weight functions computed with direct diagonalisation around $\ww=0$ is relatively high. The reason for this is that close to $\ww=0$ the eigenfrequency distribution rapidly drops to zero, which means that there are only a few modes present in the vicinity of $\ww=0$. In the KPM, the $\delta$-peaks which constitute the spectrum $\rho(\ww)$ are approximated by a distribution of finite width~\cite{Beltukov2016}. As explained in Section A of this Appendix, the resolution capability of the approximation is set by the maximum degree of the Chebyshev polynomials used in the truncated series expansion of $\rho(\ww)$. Hence, a correct accounting of the position and relative frequency of the very low-lying eigenfrequencies requires high-degree polynomials.

\begin{figure}
\includegraphics[width=8cm]{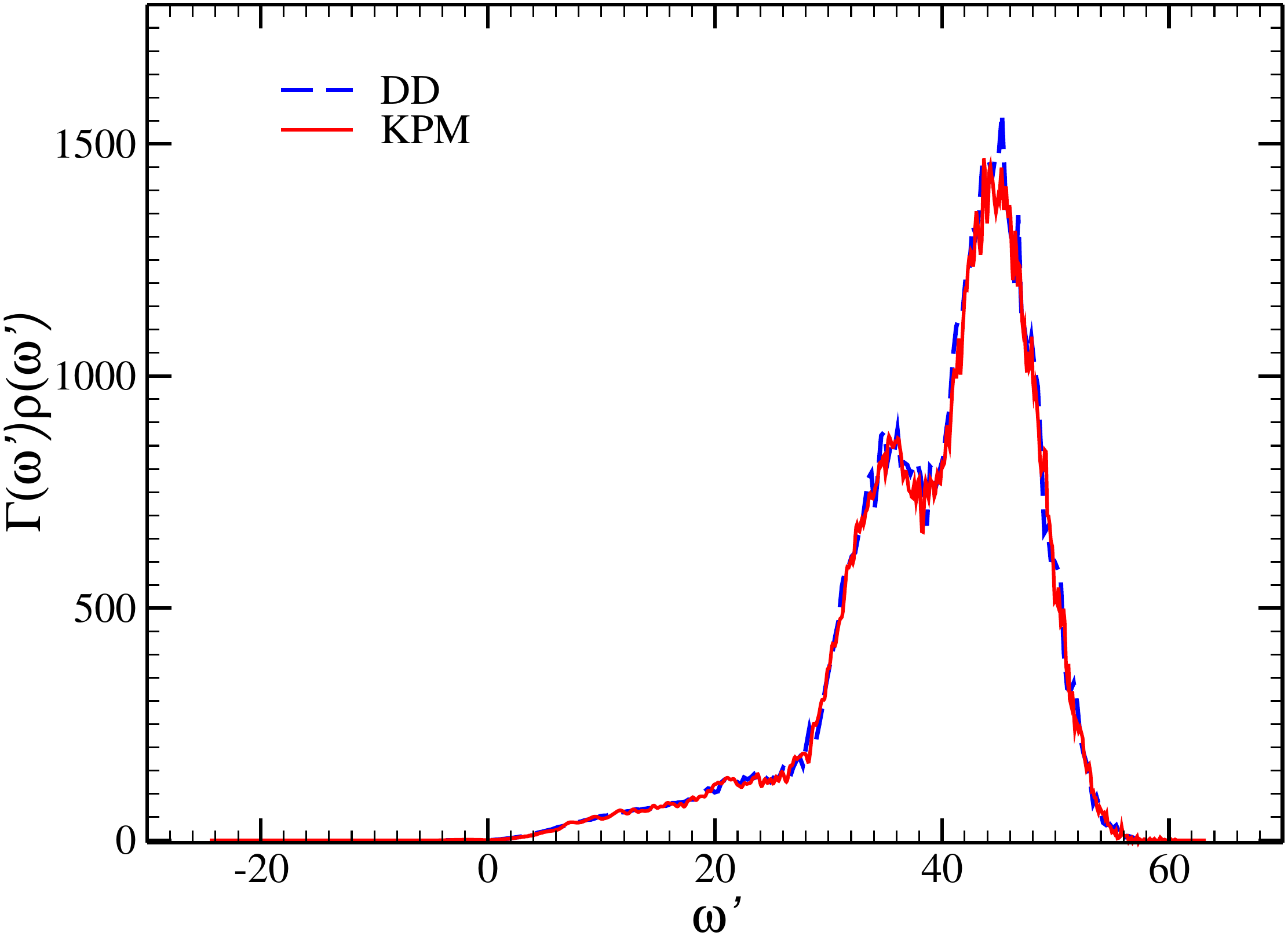}
\caption{The product of the DOS $\rho(\ww)$ and the non-affine correlator $\Gamma(\ww)$ comparing the results from direct diagonalisation with KPM.}
\label{fig:3mass_GammaDOS}
\end{figure}

We can observe in Fig.~\ref{fig:weightF_plot} that at zero frequency the contributions from masses $m_1$ and $m_2$ are equal, i.e. $\chi_1(0)=\chi_2(0)  \approx 0.5$. This value reflects the fraction of particles of different masses in the polymer chains, because the zero-frequency mode corresponds to a global translation of the system. The contribution to the corresponding zero-frequency eigenvector therefore has to be equal for every single particle, since this global zero-frequency displacement is independent of the mass of the particle. This leads to the zero-frequency eigenvector $\gEigV(\wwp=0)$ reflecting the relative fractions of different mass species~\cite{Taraskin1997}.


At low frequencies the weight functions of species 1 ($m_1$ = 1) and species 2 ($m_2$ = 3) yield about the same contribution, and cross over at a frequency which roughly corresponds to the first large Lennard-Jones peak of the VDOS. At higher frequencies, the weight function of species 1, $\chi_1(\ww)$, increases and saturates while for $\chi_2(\ww)$ it is the opposite (Fig.~\ref{fig:weightF_plot}). In the high-frequency limit, we observe that most of the contribution to the overall weight function comes from the lightest particles, species 1.

The full VDOS from KPM, depicted in Fig. \ref{fig:DOS_full}, closely matches the results obtained from direct diagonalization. Moreover, we can use the weight functions $\chi_n(\ww)$ to straightforwardly compute the partial densities of states for each species by using Eq. \ref{eq:weightDOS}. The results for the pDOS obtained from direct diagonalization and KPM also match very well (Figs. \ref{fig:pDOS_m1}-\ref{fig:pDOS_m2}). Note that the fluctuations in the double peak at high-frequency are lower in the case of the KPM result due to the much larger polymer system consisting of $N=1 \times 10^5$ particles. 

In addition to the large Lennard-Jones peak at low-frequencies, we notice that in comparison to the shape of the VDOS of the single-mass polymer system \cite{Milkus2018}, the high-frequency FENE peak has split into two smaller peaks. In Fig. \ref{fig:3mass_fullDOS} we show how the two partial densities of states sum up to the full VDOS. It is interesting to observe that this double peak is comprised almost exclusively of modes from the lightest masses $m_1=1$, likely representing fast oscillations of the $m_1$ with respect to $m_2$, which is a factor of three heavier (see also \cite{Milkus2018} for the discussion of the physical origin of the VDOS peaks).

Figure~\ref{fig:3mass_GammaDOS} shows the product $\Gamma(\ww) \rho(\ww)$, which is the direct output of the KPM algorithm, obtained for the multi-component non-affine correlator. Again the agreement between the direct diagonalization and KPM results is excellent.

Finally, we use $\Gamma(\ww) \rho(\ww)$ to calculate the components of the complex viscoelastic shear modulus obtained by different methods in Figs. 1b-c of the main text. Following \cite{ElderZacconeSirk:2019} we introduced here a low-frequency cutoff $\ww_{cut}=1$, i.e. the frequencies with absolute value $\ww\leq \ww_{cut}$ were excluded from the integration in Eq. ~(\ref{eq:Gamma_contin}). This cutoff eliminates spurious contributions from poorly sampled low-frequency regions, which otherwise lead to large uncertainties in the low-frequency modulus. The exact value can be estimated as $\ww_{min} =  \frac{c_s}{L_0}$, where $c_s$
is the shear wave sound speed and $L_0$ is the box length. It also means that since KPM give possibility to approach larger system it also able to sample lower $\ww$.
The agreement between KPM and direct diagonalisation is excellent. The MD results also show excellent agreement with the DD/KPM results for $G'$ and $G''$.

\bibliography{KPM}

\begin{thebibliography}{33}%
\makeatletter
\providecommand \@ifxundefined [1]{%
 \@ifx{#1\undefined}
}%
\providecommand \@ifnum [1]{%
 \ifnum #1\expandafter \@firstoftwo
 \else \expandafter \@secondoftwo
 \fi
}%
\providecommand \@ifx [1]{%
 \ifx #1\expandafter \@firstoftwo
 \else \expandafter \@secondoftwo
 \fi
}%
\providecommand \natexlab [1]{#1}%
\providecommand \enquote  [1]{``#1''}%
\providecommand \bibnamefont  [1]{#1}%
\providecommand \bibfnamefont [1]{#1}%
\providecommand \citenamefont [1]{#1}%
\providecommand \href@noop [0]{\@secondoftwo}%
\providecommand \href [0]{\begingroup \@sanitize@url \@href}%
\providecommand \@href[1]{\@@startlink{#1}\@@href}%
\providecommand \@@href[1]{\endgroup#1\@@endlink}%
\providecommand \@sanitize@url [0]{\catcode `\\12\catcode `\$12\catcode
  `\&12\catcode `\#12\catcode `\^12\catcode `\_12\catcode `\%12\relax}%
\providecommand \@@startlink[1]{}%
\providecommand \@@endlink[0]{}%
\providecommand \url  [0]{\begingroup\@sanitize@url \@url }%
\providecommand \@url [1]{\endgroup\@href {#1}{\urlprefix }}%
\providecommand \urlprefix  [0]{URL }%
\providecommand \Eprint [0]{\href }%
\providecommand \doibase [0]{http://dx.doi.org/}%
\providecommand \selectlanguage [0]{\@gobble}%
\providecommand \bibinfo  [0]{\@secondoftwo}%
\providecommand \bibfield  [0]{\@secondoftwo}%
\providecommand \translation [1]{[#1]}%
\providecommand \BibitemOpen [0]{}%
\providecommand \bibitemStop [0]{}%
\providecommand \bibitemNoStop [0]{.\EOS\space}%
\providecommand \EOS [0]{\spacefactor3000\relax}%
\providecommand \BibitemShut  [1]{\csname bibitem#1\endcsname}%
\let\auto@bib@innerbib\@empty
\bibitem [{\citenamefont {Born}\ and\ \citenamefont {Huang}(1954)}]{BornHuang}%
  \BibitemOpen
  \bibfield  {author} {\bibinfo {author} {\bibfnamefont {M.}~\bibnamefont
  {Born}}\ and\ \bibinfo {author} {\bibfnamefont {K.}~\bibnamefont {Huang}},\
  }\href@noop {} {\emph {\bibinfo {title} {Dynamical Theory of Crystal
  Lattices}}}\ (\bibinfo  {publisher} {Clarendon Press},\ \bibinfo {address}
  {Oxford},\ \bibinfo {year} {1954})\BibitemShut {NoStop}%
\bibitem [{\citenamefont {Lema{\^i}tre}\ and\ \citenamefont
  {Maloney}(2006)}]{Lemaitre2006}%
  \BibitemOpen
  \bibfield  {author} {\bibinfo {author} {\bibfnamefont {A.}~\bibnamefont
  {Lema{\^i}tre}}\ and\ \bibinfo {author} {\bibfnamefont {C.}~\bibnamefont
  {Maloney}},\ }\href@noop {} {\bibfield  {journal} {\bibinfo  {journal} {J.
  Stat. Phys.}\ }\textbf {\bibinfo {volume} {123}},\ \bibinfo {pages} {415}
  (\bibinfo {year} {2006})}\BibitemShut {NoStop}%
\bibitem [{\citenamefont {Milkus}\ and\ \citenamefont
  {Zaccone}(2017)}]{Milkus2017}%
  \BibitemOpen
  \bibfield  {author} {\bibinfo {author} {\bibfnamefont {R.}~\bibnamefont
  {Milkus}}\ and\ \bibinfo {author} {\bibfnamefont {A.}~\bibnamefont
  {Zaccone}},\ }\href {\doibase 10.1103/PhysRevE.95.023001} {\bibfield
  {journal} {\bibinfo  {journal} {Phys. Rev. E}\ }\textbf {\bibinfo {volume}
  {95}},\ \bibinfo {pages} {023001} (\bibinfo {year} {2017})}\BibitemShut
  {NoStop}%
\bibitem [{\citenamefont {Palyulin}\ \emph {et~al.}(2018)\citenamefont
  {Palyulin}, \citenamefont {Ness}, \citenamefont {Milkus}, \citenamefont
  {Elder}, \citenamefont {Sirk},\ and\ \citenamefont
  {Zaccone}}]{Prediction2018}%
  \BibitemOpen
  \bibfield  {author} {\bibinfo {author} {\bibfnamefont {V.~V.}\ \bibnamefont
  {Palyulin}}, \bibinfo {author} {\bibfnamefont {C.}~\bibnamefont {Ness}},
  \bibinfo {author} {\bibfnamefont {R.}~\bibnamefont {Milkus}}, \bibinfo
  {author} {\bibfnamefont {R.~M.}\ \bibnamefont {Elder}}, \bibinfo {author}
  {\bibfnamefont {T.~W.}\ \bibnamefont {Sirk}}, \ and\ \bibinfo {author}
  {\bibfnamefont {A.}~\bibnamefont {Zaccone}},\ }\href@noop {} {\bibfield
  {journal} {\bibinfo  {journal} {Soft Matter}\ }\textbf {\bibinfo {volume}
  {14}},\ \bibinfo {pages} {8475} (\bibinfo {year} {2018})}\BibitemShut
  {NoStop}%
\bibitem [{\citenamefont {Ness}\ \emph {et~al.}(2017)\citenamefont {Ness},
  \citenamefont {Palyulin}, \citenamefont {Milkus}, \citenamefont {Elder},
  \citenamefont {Sirk},\ and\ \citenamefont {Zaccone}}]{stiffness}%
  \BibitemOpen
  \bibfield  {author} {\bibinfo {author} {\bibfnamefont {C.}~\bibnamefont
  {Ness}}, \bibinfo {author} {\bibfnamefont {V.}~\bibnamefont {Palyulin}},
  \bibinfo {author} {\bibfnamefont {R.}~\bibnamefont {Milkus}}, \bibinfo
  {author} {\bibfnamefont {R.}~\bibnamefont {Elder}}, \bibinfo {author}
  {\bibfnamefont {T.}~\bibnamefont {Sirk}}, \ and\ \bibinfo {author}
  {\bibfnamefont {A.}~\bibnamefont {Zaccone}},\ }\href@noop {} {\bibfield
  {journal} {\bibinfo  {journal} {Phys. Rev. E}\ }\textbf {\bibinfo {volume}
  {96}},\ \bibinfo {pages} {030501(R)} (\bibinfo {year} {2017})}\BibitemShut
  {NoStop}%
\bibitem [{\citenamefont {Lacks}\ and\ \citenamefont
  {Rutledge}(1994)}]{Rutledge1994}%
  \BibitemOpen
  \bibfield  {author} {\bibinfo {author} {\bibfnamefont {D.~J.}\ \bibnamefont
  {Lacks}}\ and\ \bibinfo {author} {\bibfnamefont {G.~C.}\ \bibnamefont
  {Rutledge}},\ }\href@noop {} {\bibfield  {journal} {\bibinfo  {journal} {J.
  Phys. Chem.}\ }\textbf {\bibinfo {volume} {98}},\ \bibinfo {pages} {1222}
  (\bibinfo {year} {1994})}\BibitemShut {NoStop}%
\bibitem [{\citenamefont {Elder}\ \emph {et~al.}(2019)\citenamefont {Elder},
  \citenamefont {Zaccone},\ and\ \citenamefont {Sirk}}]{ElderZacconeSirk:2019}%
  \BibitemOpen
  \bibfield  {author} {\bibinfo {author} {\bibfnamefont {R.~M.}\ \bibnamefont
  {Elder}}, \bibinfo {author} {\bibfnamefont {A.}~\bibnamefont {Zaccone}}, \
  and\ \bibinfo {author} {\bibfnamefont {T.~W.}\ \bibnamefont {Sirk}},\
  }\href@noop {} {\bibfield  {journal} {\bibinfo  {journal} {ACS Macro
  Letters}\ }\textbf {\bibinfo {volume} {8}},\ \bibinfo {pages} {1160}
  (\bibinfo {year} {2019})}\BibitemShut {NoStop}%
\bibitem [{\citenamefont {Mazzacurati}\ \emph {et~al.}(1996)\citenamefont
  {Mazzacurati}, \citenamefont {Ruocco},\ and\ \citenamefont
  {Sampoli}}]{Ruocco1996}%
  \BibitemOpen
  \bibfield  {author} {\bibinfo {author} {\bibfnamefont {V.}~\bibnamefont
  {Mazzacurati}}, \bibinfo {author} {\bibfnamefont {G.}~\bibnamefont {Ruocco}},
  \ and\ \bibinfo {author} {\bibfnamefont {M.}~\bibnamefont {Sampoli}},\ }\href
  {\doibase 10.1209/epl/i1996-00515-8} {\bibfield  {journal} {\bibinfo
  {journal} {Europhysics Letters ({EPL})}\ }\textbf {\bibinfo {volume} {34}},\
  \bibinfo {pages} {681} (\bibinfo {year} {1996})}\BibitemShut {NoStop}%
\bibitem [{\citenamefont {Beltukov}\ \emph {et~al.}(2016)\citenamefont
  {Beltukov}, \citenamefont {Fusco}, \citenamefont {Parshin},\ and\
  \citenamefont {Tanguy}}]{Beltukov2016}%
  \BibitemOpen
  \bibfield  {author} {\bibinfo {author} {\bibfnamefont {Y.~M.}\ \bibnamefont
  {Beltukov}}, \bibinfo {author} {\bibfnamefont {C.}~\bibnamefont {Fusco}},
  \bibinfo {author} {\bibfnamefont {D.~A.}\ \bibnamefont {Parshin}}, \ and\
  \bibinfo {author} {\bibfnamefont {A.}~\bibnamefont {Tanguy}},\ }\href
  {\doibase 10.1103/PhysRevE.93.023006} {\bibfield  {journal} {\bibinfo
  {journal} {Phys. Rev. E}\ }\textbf {\bibinfo {volume} {93}},\ \bibinfo
  {pages} {023006} (\bibinfo {year} {2016})}\BibitemShut {NoStop}%
\bibitem [{\citenamefont {VanderWerf}\ \emph {et~al.}(2018)\citenamefont
  {VanderWerf}, \citenamefont {Jin}, \citenamefont {Shattuck},\ and\
  \citenamefont {O'Hern}}]{OHern2018}%
  \BibitemOpen
  \bibfield  {author} {\bibinfo {author} {\bibfnamefont {K.}~\bibnamefont
  {VanderWerf}}, \bibinfo {author} {\bibfnamefont {W.}~\bibnamefont {Jin}},
  \bibinfo {author} {\bibfnamefont {M.~D.}\ \bibnamefont {Shattuck}}, \ and\
  \bibinfo {author} {\bibfnamefont {C.~S.}\ \bibnamefont {O'Hern}},\ }\href
  {\doibase 10.1103/PhysRevE.97.012909} {\bibfield  {journal} {\bibinfo
  {journal} {Phys. Rev. E}\ }\textbf {\bibinfo {volume} {97}},\ \bibinfo
  {pages} {012909} (\bibinfo {year} {2018})}\BibitemShut {NoStop}%
\bibitem [{\citenamefont {Xu}\ \emph {et~al.}(2009)\citenamefont {Xu},
  \citenamefont {Vitelli}, \citenamefont {Wyart}, \citenamefont {Liu},\ and\
  \citenamefont {Nagel}}]{Vitelli2009}%
  \BibitemOpen
  \bibfield  {author} {\bibinfo {author} {\bibfnamefont {N.}~\bibnamefont
  {Xu}}, \bibinfo {author} {\bibfnamefont {V.}~\bibnamefont {Vitelli}},
  \bibinfo {author} {\bibfnamefont {M.}~\bibnamefont {Wyart}}, \bibinfo
  {author} {\bibfnamefont {A.~J.}\ \bibnamefont {Liu}}, \ and\ \bibinfo
  {author} {\bibfnamefont {S.~R.}\ \bibnamefont {Nagel}},\ }\href {\doibase
  10.1103/PhysRevLett.102.038001} {\bibfield  {journal} {\bibinfo  {journal}
  {Phys. Rev. Lett.}\ }\textbf {\bibinfo {volume} {102}},\ \bibinfo {pages}
  {038001} (\bibinfo {year} {2009})}\BibitemShut {NoStop}%
\bibitem [{\citenamefont {Tighe}(2011)}]{Tighe2011}%
  \BibitemOpen
  \bibfield  {author} {\bibinfo {author} {\bibfnamefont {B.~P.}\ \bibnamefont
  {Tighe}},\ }\href {\doibase 10.1103/PhysRevLett.107.158303} {\bibfield
  {journal} {\bibinfo  {journal} {Phys. Rev. Lett.}\ }\textbf {\bibinfo
  {volume} {107}},\ \bibinfo {pages} {158303} (\bibinfo {year}
  {2011})}\BibitemShut {NoStop}%
\bibitem [{\citenamefont {Mizuno}\ \emph {et~al.}(2016)\citenamefont {Mizuno},
  \citenamefont {Mossa},\ and\ \citenamefont {Barrat}}]{Mizuno2016}%
  \BibitemOpen
  \bibfield  {author} {\bibinfo {author} {\bibfnamefont {H.}~\bibnamefont
  {Mizuno}}, \bibinfo {author} {\bibfnamefont {S.}~\bibnamefont {Mossa}}, \
  and\ \bibinfo {author} {\bibfnamefont {J.-L.}\ \bibnamefont {Barrat}},\
  }\href {\doibase 10.1103/PhysRevB.94.144303} {\bibfield  {journal} {\bibinfo
  {journal} {Phys. Rev. B}\ }\textbf {\bibinfo {volume} {94}},\ \bibinfo
  {pages} {144303} (\bibinfo {year} {2016})}\BibitemShut {NoStop}%
\bibitem [{\citenamefont {Ikeda}\ \emph {et~al.}(2020)\citenamefont {Ikeda},
  \citenamefont {Kawasaki}, \citenamefont {Berthier}, \citenamefont {Saitoh},\
  and\ \citenamefont {Hatano}}]{Ikeda2020}%
  \BibitemOpen
  \bibfield  {author} {\bibinfo {author} {\bibfnamefont {A.}~\bibnamefont
  {Ikeda}}, \bibinfo {author} {\bibfnamefont {T.}~\bibnamefont {Kawasaki}},
  \bibinfo {author} {\bibfnamefont {L.}~\bibnamefont {Berthier}}, \bibinfo
  {author} {\bibfnamefont {K.}~\bibnamefont {Saitoh}}, \ and\ \bibinfo {author}
  {\bibfnamefont {T.}~\bibnamefont {Hatano}},\ }\href {\doibase
  10.1103/PhysRevLett.124.058001} {\bibfield  {journal} {\bibinfo  {journal}
  {Phys. Rev. Lett.}\ }\textbf {\bibinfo {volume} {124}},\ \bibinfo {pages}
  {058001} (\bibinfo {year} {2020})}\BibitemShut {NoStop}%
\bibitem [{\citenamefont {Rocklin}\ \emph {et~al.}(2016)\citenamefont
  {Rocklin}, \citenamefont {Chen}, \citenamefont {Falk}, \citenamefont
  {Vitelli},\ and\ \citenamefont {Lubensky}}]{Vitelli2016}%
  \BibitemOpen
  \bibfield  {author} {\bibinfo {author} {\bibfnamefont {D.~Z.}\ \bibnamefont
  {Rocklin}}, \bibinfo {author} {\bibfnamefont {B.~G.-g.}\ \bibnamefont
  {Chen}}, \bibinfo {author} {\bibfnamefont {M.}~\bibnamefont {Falk}}, \bibinfo
  {author} {\bibfnamefont {V.}~\bibnamefont {Vitelli}}, \ and\ \bibinfo
  {author} {\bibfnamefont {T.~C.}\ \bibnamefont {Lubensky}},\ }\href {\doibase
  10.1103/PhysRevLett.116.135503} {\bibfield  {journal} {\bibinfo  {journal}
  {Phys. Rev. Lett.}\ }\textbf {\bibinfo {volume} {116}},\ \bibinfo {pages}
  {135503} (\bibinfo {year} {2016})}\BibitemShut {NoStop}%
\bibitem [{\citenamefont {Rudyak}\ \emph {et~al.}(2017)\citenamefont {Rudyak},
  \citenamefont {Gavrilov}, \citenamefont {Guseva},\ and\ \citenamefont
  {Chertovich}}]{atomistic1}%
  \BibitemOpen
  \bibfield  {author} {\bibinfo {author} {\bibfnamefont {V.~Y.}\ \bibnamefont
  {Rudyak}}, \bibinfo {author} {\bibfnamefont {A.~A.}\ \bibnamefont
  {Gavrilov}}, \bibinfo {author} {\bibfnamefont {D.~V.}\ \bibnamefont
  {Guseva}}, \ and\ \bibinfo {author} {\bibfnamefont {A.~V.}\ \bibnamefont
  {Chertovich}},\ }\href@noop {} {\bibfield  {journal} {\bibinfo  {journal}
  {Macromolecular Theory and Simulations}\ }\textbf {\bibinfo {volume} {26}},\
  \bibinfo {pages} {1700015} (\bibinfo {year} {2017})}\BibitemShut {NoStop}%
\bibitem [{\citenamefont {Guseva}\ \emph {et~al.}(2018)\citenamefont {Guseva},
  \citenamefont {Rudyak}, \citenamefont {Komarov}, \citenamefont {Sulimov},
  \citenamefont {Bulgakov},\ and\ \citenamefont {Chertovich}}]{atomistic2}%
  \BibitemOpen
  \bibfield  {author} {\bibinfo {author} {\bibfnamefont {D.}~\bibnamefont
  {Guseva}}, \bibinfo {author} {\bibfnamefont {V.}~\bibnamefont {Rudyak}},
  \bibinfo {author} {\bibfnamefont {P.}~\bibnamefont {Komarov}}, \bibinfo
  {author} {\bibfnamefont {A.}~\bibnamefont {Sulimov}}, \bibinfo {author}
  {\bibfnamefont {B.}~\bibnamefont {Bulgakov}}, \ and\ \bibinfo {author}
  {\bibfnamefont {A.}~\bibnamefont {Chertovich}},\ }\href@noop {} {\bibfield
  {journal} {\bibinfo  {journal} {J. Polym. Sci., Part B: Polym. Phys.}\
  }\textbf {\bibinfo {volume} {56}},\ \bibinfo {pages} {362} (\bibinfo {year}
  {2018})}\BibitemShut {NoStop}%
\bibitem [{\citenamefont {Damart}\ \emph {et~al.}(2017)\citenamefont {Damart},
  \citenamefont {Tanguy},\ and\ \citenamefont {Rodney}}]{Rodney2017}%
  \BibitemOpen
  \bibfield  {author} {\bibinfo {author} {\bibfnamefont {T.}~\bibnamefont
  {Damart}}, \bibinfo {author} {\bibfnamefont {A.}~\bibnamefont {Tanguy}}, \
  and\ \bibinfo {author} {\bibfnamefont {D.}~\bibnamefont {Rodney}},\
  }\href@noop {} {\bibfield  {journal} {\bibinfo  {journal} {Phys. Rev. B}\
  }\textbf {\bibinfo {volume} {95(5)}},\ \bibinfo {pages} {054203} (\bibinfo
  {year} {2017})}\BibitemShut {NoStop}%
\bibitem [{\citenamefont {Kremer}\ and\ \citenamefont
  {Grest}(1986)}]{Kremer1986}%
  \BibitemOpen
  \bibfield  {author} {\bibinfo {author} {\bibfnamefont {K.}~\bibnamefont
  {Kremer}}\ and\ \bibinfo {author} {\bibfnamefont {G.~S.}\ \bibnamefont
  {Grest}},\ }\href@noop {} {\bibfield  {journal} {\bibinfo  {journal} {Phys.
  Rev. A}\ }\textbf {\bibinfo {volume} {33}},\ \bibinfo {pages} {3628}
  (\bibinfo {year} {1986})}\BibitemShut {NoStop}%
\bibitem [{Sup()}]{Suppl}%
  \BibitemOpen
  \href@noop {} {\enquote {\bibinfo {title} {Supplementary information
  available at...}}\ }\BibitemShut {NoStop}%
\bibitem [{\citenamefont {Rahman}\ \emph {et~al.}(1976)\citenamefont {Rahman},
  \citenamefont {Mandell},\ and\ \citenamefont {McTague}}]{Rahman1976}%
  \BibitemOpen
  \bibfield  {author} {\bibinfo {author} {\bibfnamefont {A.}~\bibnamefont
  {Rahman}}, \bibinfo {author} {\bibfnamefont {M.}~\bibnamefont {Mandell}}, \
  and\ \bibinfo {author} {\bibfnamefont {J.}~\bibnamefont {McTague}},\
  }\href@noop {} {\bibfield  {journal} {\bibinfo  {journal} {J. Chem. Phys.}\
  }\textbf {\bibinfo {volume} {64}},\ \bibinfo {pages} {1564} (\bibinfo {year}
  {1976})}\BibitemShut {NoStop}%
\bibitem [{\citenamefont {Plimpton}(1995)}]{LAMMPS}%
  \BibitemOpen
  \bibfield  {author} {\bibinfo {author} {\bibfnamefont {S.}~\bibnamefont
  {Plimpton}},\ }\href@noop {} {\bibfield  {journal} {\bibinfo  {journal} {J.
  Comp. Phys}\ }\textbf {\bibinfo {volume} {117}},\ \bibinfo {pages} {1}
  (\bibinfo {year} {1995})},\ \bibinfo {note} {see also:
  http://lammps.sandia.gov}\BibitemShut {NoStop}%
\bibitem [{\citenamefont {Milkus}\ \emph {et~al.}(2018)\citenamefont {Milkus},
  \citenamefont {Ness}, \citenamefont {Palyulin}, \citenamefont {Weber},
  \citenamefont {Lapkin},\ and\ \citenamefont {Zaccone}}]{Milkus2018}%
  \BibitemOpen
  \bibfield  {author} {\bibinfo {author} {\bibfnamefont {R.}~\bibnamefont
  {Milkus}}, \bibinfo {author} {\bibfnamefont {C.}~\bibnamefont {Ness}},
  \bibinfo {author} {\bibfnamefont {V.~V.}\ \bibnamefont {Palyulin}}, \bibinfo
  {author} {\bibfnamefont {J.}~\bibnamefont {Weber}}, \bibinfo {author}
  {\bibfnamefont {A.}~\bibnamefont {Lapkin}}, \ and\ \bibinfo {author}
  {\bibfnamefont {A.}~\bibnamefont {Zaccone}},\ }\href@noop {} {\bibfield
  {journal} {\bibinfo  {journal} {Macromolecules}\ }\textbf {\bibinfo {volume}
  {51}},\ \bibinfo {pages} {1559} (\bibinfo {year} {2018})}\BibitemShut
  {NoStop}%
\bibitem [{\citenamefont {Taraskin}\ and\ \citenamefont
  {Elliott}(1997)}]{Taraskin1997}%
  \BibitemOpen
  \bibfield  {author} {\bibinfo {author} {\bibfnamefont {S.~N.}\ \bibnamefont
  {Taraskin}}\ and\ \bibinfo {author} {\bibfnamefont {S.~R.}\ \bibnamefont
  {Elliott}},\ }\href@noop {} {\bibfield  {journal} {\bibinfo  {journal} {Phys.
  Rev. B}\ }\textbf {\bibinfo {volume} {55}},\ \bibinfo {pages} {117} (\bibinfo
  {year} {1997})}\BibitemShut {NoStop}%
\bibitem [{\citenamefont {Zaccone}\ and\ \citenamefont
  {Scossa-Romano}(2011)}]{Zaccone2011}%
  \BibitemOpen
  \bibfield  {author} {\bibinfo {author} {\bibfnamefont {A.}~\bibnamefont
  {Zaccone}}\ and\ \bibinfo {author} {\bibfnamefont {E.}~\bibnamefont
  {Scossa-Romano}},\ }\href {\doibase 10.1103/PhysRevB.83.184205} {\bibfield
  {journal} {\bibinfo  {journal} {Phys. Rev. B}\ }\textbf {\bibinfo {volume}
  {83}},\ \bibinfo {pages} {184205} (\bibinfo {year} {2011})}\BibitemShut
  {NoStop}%
\bibitem [{\citenamefont {Krausser}(8051)}]{JohannesThesis}%
  \BibitemOpen
  \bibfield  {author} {\bibinfo {author} {\bibfnamefont {J.}~\bibnamefont
  {Krausser}},\ }\href@noop {} {\emph {\bibinfo {title} {Non-affine lattice
  dynamics of disordered solids (Doctoral thesis under the supervision of A.
  Zaccone).}}}\ (\bibinfo {year} {2018, DOI:
  https://doi.org/10.17863/CAM.28051})\BibitemShut {NoStop}%
\bibitem [{\citenamefont {Veseli{\'{c}}}(2011)}]{Veselic2011}%
  \BibitemOpen
  \bibfield  {author} {\bibinfo {author} {\bibfnamefont {K.}~\bibnamefont
  {Veseli{\'{c}}}},\ }\href@noop {} {\emph {\bibinfo {title} {Damped
  oscillations of linear systems: A mathematical introduction, Lecture Notes in
  Mathematics}}},\ Vol.\ \bibinfo {volume} {2023}\ (\bibinfo  {publisher}
  {Springer},\ \bibinfo {year} {2011})\ pp.\ \bibinfo {pages}
  {1--226}\BibitemShut {NoStop}%
\bibitem [{\citenamefont {Stratt}(1995)}]{Stratt1995}%
  \BibitemOpen
  \bibfield  {author} {\bibinfo {author} {\bibfnamefont {R.}~\bibnamefont
  {Stratt}},\ }\href@noop {} {\bibfield  {journal} {\bibinfo  {journal}
  {Macromolecular Theory and Simulations}\ }\textbf {\bibinfo {volume} {28}},\
  \bibinfo {pages} {201–207} (\bibinfo {year} {1995})}\BibitemShut {NoStop}%
\bibitem [{\citenamefont {Keyes}(1997)}]{Keyes1997}%
  \BibitemOpen
  \bibfield  {author} {\bibinfo {author} {\bibfnamefont {T.}~\bibnamefont
  {Keyes}},\ }\href@noop {} {\bibfield  {journal} {\bibinfo  {journal} {J.
  Phys. Chem. A}\ }\textbf {\bibinfo {volume} {101}},\ \bibinfo {pages} {2921}
  (\bibinfo {year} {1997})}\BibitemShut {NoStop}%
\bibitem [{\citenamefont {Zhang}\ \emph {et~al.}(2019)\citenamefont {Zhang},
  \citenamefont {Douglas},\ and\ \citenamefont {Starr}}]{Douglas2019}%
  \BibitemOpen
  \bibfield  {author} {\bibinfo {author} {\bibfnamefont {W.}~\bibnamefont
  {Zhang}}, \bibinfo {author} {\bibfnamefont {J.~F.}\ \bibnamefont {Douglas}},
  \ and\ \bibinfo {author} {\bibfnamefont {F.~W.}\ \bibnamefont {Starr}},\
  }\href {\doibase 10.1063/1.5127821} {\bibfield  {journal} {\bibinfo
  {journal} {The Journal of Chemical Physics}\ }\textbf {\bibinfo {volume}
  {151}},\ \bibinfo {pages} {184904} (\bibinfo {year} {2019})}\BibitemShut
  {NoStop}%
\bibitem [{\citenamefont {Wei{\ss}e}\ \emph {et~al.}(2016)\citenamefont
  {Wei{\ss}e}, \citenamefont {Wellein}, \citenamefont {Alvermann},\ and\
  \citenamefont {Fehske}}]{Weisse2006}%
  \BibitemOpen
  \bibfield  {author} {\bibinfo {author} {\bibfnamefont {A.}~\bibnamefont
  {Wei{\ss}e}}, \bibinfo {author} {\bibfnamefont {G.}~\bibnamefont {Wellein}},
  \bibinfo {author} {\bibfnamefont {A.}~\bibnamefont {Alvermann}}, \ and\
  \bibinfo {author} {\bibfnamefont {H.}~\bibnamefont {Fehske}},\ }\href@noop {}
  {\bibfield  {journal} {\bibinfo  {journal} {Rev. Mod. Phys.}\ }\textbf
  {\bibinfo {volume} {78}},\ \bibinfo {pages} {275} (\bibinfo {year}
  {2016})}\BibitemShut {NoStop}%
\bibitem [{\citenamefont {Cui}\ \emph {et~al.}(2019)\citenamefont {Cui},
  \citenamefont {Zaccone},\ and\ \citenamefont {Rodney}}]{Cui2019}%
  \BibitemOpen
  \bibfield  {author} {\bibinfo {author} {\bibfnamefont {B.}~\bibnamefont
  {Cui}}, \bibinfo {author} {\bibfnamefont {A.}~\bibnamefont {Zaccone}}, \ and\
  \bibinfo {author} {\bibfnamefont {D.}~\bibnamefont {Rodney}},\ }\href
  {\doibase 10.1063/1.5129025} {\bibfield  {journal} {\bibinfo  {journal} {J.
  Chem. Phys.}\ }\textbf {\bibinfo {volume} {151}},\ \bibinfo {pages} {224509}
  (\bibinfo {year} {2019})}\BibitemShut {NoStop}%
\bibitem [{\citenamefont {Shenoy}\ \emph {et~al.}(1999)\citenamefont {Shenoy},
  \citenamefont {Miller}, \citenamefont {Tadmor}, \citenamefont {Rodney},
  \citenamefont {Phillips},\ and\ \citenamefont {Ortiz}}]{Rodney1999}%
  \BibitemOpen
  \bibfield  {author} {\bibinfo {author} {\bibfnamefont {V.}~\bibnamefont
  {Shenoy}}, \bibinfo {author} {\bibfnamefont {R.}~\bibnamefont {Miller}},
  \bibinfo {author} {\bibfnamefont {E.}~\bibnamefont {Tadmor}}, \bibinfo
  {author} {\bibfnamefont {D.}~\bibnamefont {Rodney}}, \bibinfo {author}
  {\bibfnamefont {R.}~\bibnamefont {Phillips}}, \ and\ \bibinfo {author}
  {\bibfnamefont {M.}~\bibnamefont {Ortiz}},\ }\href {\doibase
  https://doi.org/10.1016/S0022-5096(98)00051-9} {\bibfield  {journal}
  {\bibinfo  {journal} {J. Mech. Phys. Solids}\ }\textbf {\bibinfo {volume}
  {47}},\ \bibinfo {pages} {611 } (\bibinfo {year} {1999})}\BibitemShut
  {NoStop}%
\end{thebibliography}%

\end{document}